\documentclass[twocolumn]{aastex631}

\usepackage{natbib}
\usepackage{multirow}
\usepackage{booktabs}
\usepackage{epsfig}
\usepackage{ulem}
\usepackage{rotating}
\usepackage{graphicx}
\usepackage{url}
\usepackage{amsmath}
\usepackage{color}
\usepackage{savesym}
\savesymbol{tablenum}
\usepackage{siunitx}
\usepackage[normalem]{ulem}
\usepackage{appendix}
\restoresymbol{SIX}{tablenum}
\DeclareSIUnit\parsec{pc}
\DeclareSIUnit\h{\textit{h}}

\DeclareRobustCommand{\uvec}[1]{{%
  \ifcsname uvec#1\endcsname
     \csname uvec#1\endcsname
   \else
    \bm{\hat{\mathbf{#1}}}%
   \fi
}}

\newcommand{\NSNeincoherents}{584}
\newcommand{\NSNeingroups}{174}
\newcommand{\NSNenotingroups}{410}

\newcommand{\NSNeintully}{149}
\newcommand{\NSNeinlim}{121}
\newcommand{\NSNeinlam}{124}
\newcommand{\NSNeincrook}{120}

\newcommand{\chaevenautscrit}{5.0}

\newcommand{\coherentsigint}{0.140}
\newcommand{\groupsigint}{0.140}

\newcommand{\gTgrchi}{160.9}
\newcommand{\gTbrchi}{218.9}
\newcommand{\gTgmwchi}{161.2}

\newcommand{\gLimgrchi}{189.9}
\newcommand{\gLimbrchi}{217.2}
\newcommand{\gLimgmwchi}{190.0}

\newcommand{\gLamgrchi}{181.5}

\newcommand{\gCgrchi}{194.6}
\newcommand{\gchi}{253.4}

\newcommand{\gTgrnonechi}{181.4}
\newcommand{\gTbrnonechi}{237.0}
\newcommand{\gTgmwnonechi}{181.2}

\newcommand{\gLimgrnonechi}{189.0}
\newcommand{\gLimbrnonechi}{203.0}
\newcommand{\gLimgmwnonechi}{188.8}

\newcommand{\gLamgrnonechi}{193.6}

\newcommand{\gCgrnonechi}{214.3}
\newcommand{\gnonechi}{220.7}

\newcommand{\gTgrRSD}{0.178}
\newcommand{\gTbrRSD}{0.189}
\newcommand{\gTgmwRSD}{0.178}

\newcommand{\gLimgrRSD}{0.186}
\newcommand{\gLimbrRSD}{0.202}
\newcommand{\gLimgmwRSD}{0.185}

\newcommand{\gLamgrRSD}{0.189}

\newcommand{\gCgrRSD}{0.197}
\newcommand{\gRSD}{0.193}

\newcommand{\gTgrnoneRSD}{0.199}
\newcommand{\gTbrnoneRSD}{0.208}
\newcommand{\gTgmwnoneRSD}{0.199}

\newcommand{\gLimgrnoneRSD}{0.183}
\newcommand{\gLimbrnoneRSD}{0.177}
\newcommand{\gLimgmwnoneRSD}{0.184}

\newcommand{\gLamgrnoneRSD}{0.178}

\newcommand{\gCgrnoneRSD}{0.175}
\newcommand{\gnoneRSD}{0.182}

\newcommand{\gTgrmu}{$-0.031$}
\newcommand{\gTbrmu}{$-0.025$}
\newcommand{\gTgmwmu}{$-0.031$}

\newcommand{\gLimgrmu}{$-0.028$}
\newcommand{\gLimbrmu}{$-0.035$}
\newcommand{\gLimgmwmu}{$-0.028$}

\newcommand{\gLamgrmu}{$-0.033$}

\newcommand{\gCgrmu}{$-0.025$}
\newcommand{\gmu}{$-0.028$}

\newcommand{\gTgrnonemu}{$+0.002$}
\newcommand{\gTbrnonemu}{$+0.007$}
\newcommand{\gTgmwnonemu}{$+0.002$}

\newcommand{\gLimgrnonemu}{$+0.003$}
\newcommand{\gLimbrnonemu}{$-0.006$}
\newcommand{\gLimgmwnonemu}{$+0.003$}

\newcommand{\gLamgrnonemu}{$-0.001$}

\newcommand{\gCgrnonemu}{$+0.007$}
\newcommand{\gnonemu}{$0.000$}

\newcommand{\gTgrw}{$-0.09$}
\newcommand{\gTbrw}{$-0.08$}
\newcommand{\gTgmww}{$-0.09$}

\newcommand{\gLimgrw}{$-0.08$}
\newcommand{\gLimbrw}{$-0.08$}
\newcommand{\gLimgmww}{$-0.08$}

\newcommand{\gLamgrw}{$-0.09$}

\newcommand{\gCgrw}{$-0.08$}
\newcommand{\gw}{$-0.08$}

\newcommand{\gTgrnonew}{$-0.01$}
\newcommand{\gTbrnonew}{$+0.00$}
\newcommand{\gTgmwnonew}{$-0.01$}

\newcommand{\gLimgrnonew}{$+0.00$}
\newcommand{\gLimbrnonew}{$-0.00$}
\newcommand{\gLimgmwnonew}{$-0.00$}

\newcommand{\gLamgrnonew}{$-0.01$}

\newcommand{\gCgrnonew}{$+0.00$}
\newcommand{\gnonew}{$0.00$}

\newcommand{\gTgrH}{$+0.56$}
\newcommand{\gTbrH}{$+0.57$}
\newcommand{\gTgmwH}{$+0.56$}

\newcommand{\gLimgrH}{$+0.58$}
\newcommand{\gLimbrH}{$+0.60$}
\newcommand{\gLimgmwH}{$+0.58$}

\newcommand{\gLamgrH}{$+0.62$}

\newcommand{\gCgrH}{$+0.57$}
\newcommand{\gH}{$+0.60$}

\newcommand{\gTgrnoneH}{$-0.03$}
\newcommand{\gTbrnoneH}{$-0.02$}
\newcommand{\gTgmwnoneH}{$-0.03$}

\newcommand{\gLimgrnoneH}{$-0.01$}
\newcommand{\gLimbrnoneH}{$+0.00$}
\newcommand{\gLimgmwnoneH}{$-0.01$}

\newcommand{\gLamgrnoneH}{$+0.02$}

\newcommand{\gCgrnoneH}{$-0.02$}
\newcommand{\gnoneH}{$0.00$}

\newcommand{\bnonechi}{615.7}

\newcommand{\bfullchi}{629.6}

\newcommand{\bnegfullchi}{868.4}
\newcommand{\bcfchi}{688.7}

\newcommand{\bnegcfchi}{945.7}
\newcommand{\bnechi}{690.0}

\newcommand{\bvgaschi}{639.7}
\newcommand{\bnegvgaschi}{861.3}
\newcommand{\bgrchi}{583.1}
\newcommand{\bdarcychi}{563.6}
\newcommand{\bsdsschi}{586.8}
\newcommand{\bsixdfchi}{595.8}
\newcommand{\btmrschi}{582.7}
\newcommand{\bsdssdarcychi}{564.4}
\newcommand{\btmrsiloschi}{584.5}
\newcommand{\bTnonechi}{577.7}

\newcommand{\bTfullchi}{536.2}
\newcommand{\bTcfchi}{611.6}

\newcommand{\bTnechi}{593.1}
\newcommand{\bTvgaschi}{604.8}
\newcommand{\bTgrchi}{540.6}
\newcommand{\bTdarcychi}{510.3}
\newcommand{\bTsdsschi}{509.3}
\newcommand{\bTsixdfchi}{514.9}
\newcommand{\bTtmrschi}{507.9}
\newcommand{\bTsdssdarcychi}{497.2}
\newcommand{\bTtmrsiloschi}{520.3}

\newcommand{\bnoneRSD}{0.167}

\newcommand{\bfullRSD}{0.162}

\newcommand{\bnegfullRSD}{0.206}
\newcommand{\bcfRSD}{0.177}

\newcommand{\bnegcfRSD}{0.217}
\newcommand{\bneRSD}{0.165}

\newcommand{\bvgasRSD}{0.176}
\newcommand{\bnegvgasRSD}{0.186}
\newcommand{\bgrRSD}{0.156}
\newcommand{\bdarcyRSD}{0.152}
\newcommand{\bsdssRSD}{0.149}
\newcommand{\bsixdfRSD}{0.151}
\newcommand{\btmrsRSD}{0.147}
\newcommand{\bsdssdarcyRSD}{0.156}
\newcommand{\btmrsilosRSD}{0.155}
\newcommand{\bTnoneRSD}{0.170}

\newcommand{\bTfullRSD}{0.157}
\newcommand{\bTcfRSD}{0.173}

\newcommand{\bTneRSD}{0.167}
\newcommand{\bTvgasRSD}{0.177}
\newcommand{\bTgrRSD}{0.159}
\newcommand{\bTdarcyRSD}{0.149}
\newcommand{\bTsdssRSD}{0.151}
\newcommand{\bTsixdfRSD}{0.152}
\newcommand{\bTtmrsRSD}{0.149}
\newcommand{\bTsdssdarcyRSD}{0.151}
\newcommand{\bTtmrsilosRSD}{0.153}

\newcommand{\bnonemu}{$0.000$}

\newcommand{\bfullmu}{$-0.028$}

\newcommand{\bnegfullmu}{$+0.032$}
\newcommand{\bcfmu}{$-0.066$}

\newcommand{\bnegcfmu}{$+0.073$}
\newcommand{\bnemu}{$-0.019$}

\newcommand{\bvgasmu}{$-0.044$}
\newcommand{\bnegvgasmu}{$+0.048$}
\newcommand{\bgrmu}{$-0.027$}
\newcommand{\bdarcymu}{$-0.027$}
\newcommand{\bsdssmu}{$-0.024$}
\newcommand{\bsixdfmu}{$-0.025$}
\newcommand{\btmrsmu}{$-0.017$}
\newcommand{\bsdssdarcymu}{$-0.029$}
\newcommand{\btmrsilosmu}{$-0.028$}
\newcommand{\bTnonemu}{$+0.001$}

\newcommand{\bTfullmu}{$-0.029$}
\newcommand{\bTcfmu}{$-0.066$}

\newcommand{\bTnemu}{$-0.019$}
\newcommand{\bTvgasmu}{$-0.043$}
\newcommand{\bTgrmu}{$-0.026$}
\newcommand{\bTdarcymu}{$-0.027$}
\newcommand{\bTsdssmu}{$-0.024$}
\newcommand{\bTsixdfmu}{$-0.025$}
\newcommand{\bTtmrsmu}{$-0.016$}
\newcommand{\bTsdssdarcymu}{$-0.029$}
\newcommand{\bTtmrsilosmu}{$-0.027$}

\newcommand{\bnonew}{$0.00$}

\newcommand{\bfullw}{$-0.08$}

\newcommand{\bnegfullw}{$+0.08$}
\newcommand{\bcfw}{$-0.23$}

\newcommand{\bnegcfw}{$+0.18$}
\newcommand{\bnew}{$-0.05$}

\newcommand{\bvgasw}{$-0.14$}
\newcommand{\bnegvgasw}{$+0.12$}
\newcommand{\bgrw}{$-0.08$}
\newcommand{\bdarcyw}{$-0.08$}
\newcommand{\bsdssw}{$-0.07$}
\newcommand{\bsixdfw}{$-0.07$}
\newcommand{\btmrsw}{$-0.04$}
\newcommand{\bsdssdarcyw}{$-0.08$}
\newcommand{\btmrsilosw}{$-0.07$}
\newcommand{\bTnonew}{$-0.01$}

\newcommand{\bTfullw}{$-0.09$}
\newcommand{\bTcfw}{$-0.24$}

\newcommand{\bTnew}{$-0.06$}
\newcommand{\bTvgasw}{$-0.15$}
\newcommand{\bTgrw}{$-0.09$}
\newcommand{\bTdarcyw}{$-0.08$}
\newcommand{\bTsdssw}{$-0.07$}
\newcommand{\bTsixdfw}{$-0.08$}
\newcommand{\bTtmrsw}{$-0.05$}
\newcommand{\bTsdssdarcyw}{$-0.09$}
\newcommand{\bTtmrsilosw}{$-0.08$}

\newcommand{\bnoneH}{$0.00$}

\newcommand{\bfullH}{$+0.50$}

\newcommand{\bnegfullH}{$-0.61$}
\newcommand{\bcfH}{$+1.45$}

\newcommand{\bnegcfH}{$-1.64$}
\newcommand{\bneH}{$+0.40$}

\newcommand{\bvgasH}{$+0.38$}
\newcommand{\bnegvgasH}{$-0.50$}
\newcommand{\bgrH}{$+0.44$}
\newcommand{\bdarcyH}{$+0.44$}
\newcommand{\bsdssH}{$+0.37$}
\newcommand{\bsixdfH}{$+0.40$}
\newcommand{\btmrsH}{$+0.32$}
\newcommand{\bsdssdarcyH}{$+0.50$}
\newcommand{\btmrsilosH}{$+0.42$}
\newcommand{\bTnoneH}{$-0.19$}

\newcommand{\bTfullH}{$+0.35$}
\newcommand{\bTcfH}{$+1.30$}

\newcommand{\bTneH}{$+0.25$}
\newcommand{\bTvgasH}{$+0.19$}
\newcommand{\bTgrH}{$+0.27$}
\newcommand{\bTdarcyH}{$+0.27$}
\newcommand{\bTsdssH}{$+0.22$}
\newcommand{\bTsixdfH}{$+0.24$}
\newcommand{\bTtmrsH}{$+0.16$}
\newcommand{\bTsdssdarcyH}{$+0.34$}
\newcommand{\bTtmrsilosH}{$+0.25$}

\newcommand{\grindivnonechi}{220.7}
\newcommand{\grindivchi}{253.4}
\newcommand{\grTnonechi}{181.4}
\newcommand{\grTchi}{160.9}
\newcommand{\grindivnonedelchi}{0.0}
\newcommand{\grindivdelchi}{$+32.7$}
\newcommand{\grTnonedelchi}{$-39.3$}
\newcommand{\grTdelchi}{$-59.8$}
\newcommand{\grindivnoneredchi}{1.27}
\newcommand{\grindivredchi}{1.46}
\newcommand{\grTnoneredchi}{1.04}
\newcommand{\grTredchi}{0.92}
\newcommand{\grindivnonersd}{0.182}
\newcommand{\grindivrsd}{0.193}
\newcommand{\grTnonersd}{0.199}
\newcommand{\grTrsd}{0.178}

\newcommand{\ngrindivnonechi}{395.0}
\newcommand{\ngrindivchi}{376.2}
\newcommand{\ngrindivnonedelchi}{0.0}
\newcommand{\ngrindivdelchi}{$-18.8$}
\newcommand{\ngrindivnoneredchi}{0.96}
\newcommand{\ngrindivredchi}{0.92}
\newcommand{\ngrindivnonersd}{0.160}
\newcommand{\ngrindivrsd}{0.145}

\begin{document}
\title{The Pantheon+ Analysis: Evaluating Peculiar Velocity Corrections in Cosmological Analyses with Nearby Type Ia Supernovae}
    \author[0000-0001-8596-4746]{Erik R. Peterson}
    \affiliation{Department of Physics, Duke University, Durham, NC 27708, USA}
    \author[0000-0002-5153-5983]{W. D'Arcy Kenworthy}
    \affiliation{Department of Physics and Astronomy, Johns Hopkins University, Baltimore, MD 21218, USA}
    \author[0000-0002-4934-5849]{Daniel Scolnic}
    \affiliation{Department of Physics, Duke University, Durham, NC 27708, USA}
    \author{Adam G. Riess}
    \affiliation{Department of Physics and Astronomy, Johns Hopkins University, Baltimore, MD 21218, USA}
    \affiliation{Space Telescope Science Institute, Baltimore, MD 21218, USA}
    \author[0000-0001-5201-8374]{Dillon Brout}
    \affiliation{Center for Astrophysics, Harvard \& Smithsonian, 60 Garden Street, Cambridge, MA 02138, USA}
    \affiliation{NASA Einstein Fellow}
    \author[0000-0003-4074-5659]{Anthony Carr}
    \affiliation{School of Mathematics and Physics, University of Queensland, Brisbane, QLD 4072, Australia}
    \author[0000-0003-0509-1776]{H\'{e}l\`{e}ne Courtois}
    \affiliation{University Lyon 1, IUF, IP2I Lyon, 69622 Villeurbanne cedex, France}
    \author[0000-0002-4213-8783]{Tamara Davis}
    \affiliation{School of Mathematics and Physics, University of Queensland, Brisbane, QLD 4072, Australia}
    \author{Arianna Dwomoh}
    \affiliation{Department of Physics, Duke University, Durham, NC 27708, USA}
    \author[0000-0002-6230-0151]{David O. Jones}
    \affiliation{Department of Astronomy and Astrophysics, University of California, Santa Cruz, CA 92064, USA}
    \affiliation{NASA Einstein Fellow}
    \author[0000-0002-8012-6978]{Brodie Popovic}
    \affiliation{Department of Physics, Duke University, Durham, NC 27708, USA}
    \author[0000-0002-1873-8973]{Benjamin M. Rose}
    \affiliation{Department of Physics, Duke University, Durham, NC 27708, USA}
    \author{Khaled Said}
    \affiliation{School of Mathematics and Physics, University of Queensland, Brisbane, QLD 4072, Australia}

\begin{abstract}

Separating the components of redshift due to expansion and peculiar motion in the nearby universe ($z<0.1$) is critical for using Type Ia Supernovae (SNe Ia) to measure the Hubble constant ($H_0$) and the equation-of-state parameter of dark energy ($w$). Here, we study the two dominant `motions' contributing to nearby peculiar velocities: large-scale, coherent-flow (CF) motions and small-scale motions due to gravitationally associated galaxies deemed to be in a galaxy group. We use a set of $\NSNeincoherents$ low-$z$ SNe from the Pantheon+ sample, and evaluate the efficacy of corrections to these motions by measuring the improvement of SN distance residuals. 
We study multiple methods for modeling the large and small-scale motions and show that, while group assignments and CF corrections individually contribute to small improvements in Hubble residual scatter, the greatest improvement comes from the combination of the two (relative standard deviation of the Hubble residuals, Rel.~SD, improves from $\bnoneRSD$ to $\bTfullRSD$ mag). 
We find the optimal flow corrections derived from various local density maps significantly reduce Hubble residuals while raising $H_0$ by $\sim0.4$ km~s$^{-1}$ Mpc$^{-1}$ as compared to using CMB redshifts, disfavoring the hypothesis that unrecognized local structure could resolve the Hubble tension.
We estimate that the systematic uncertainties in cosmological parameters after optimally correcting redshifts are 0.06--0.11 km~s$^{-1}$ Mpc$^{-1}$ in $H_0$ and 0.02--0.03 in $w$
which are smaller than the statistical uncertainties for these measurements: 1.5 km~s$^{-1}$ Mpc$^{-1}$ for $H_0$ and 0.04 for $w$.

\end{abstract}

\section{Introduction}

Type Ia Supernovae (SNe Ia) are critical tools for measuring cosmological parameters including the Hubble constant $H_0$, which parameterizes the local expansion rate of the universe \citep{Riess16,Freedman19}, and the equation-of-state of dark energy parameter $w$ (\citealp[JLA: ][]{Betoule14}; \citealp[Pantheon: ][]{Scolnic18}; \citealp[DES: ][]{Brout19}).
Measurements of $H_0$ and/or $w$ that make use of SNe at low redshift ($0.01<z<0.1$) are sensitive to the `peculiar velocities' of host galaxies. 
Peculiar velocities (PVs) are 
the physical motion of a galaxy relative to the cosmological rest frame.

Recession velocities, $\textit{v}_\textrm{cosmo}$, are related to cosmological redshifts $\textit{z}_\textrm{cosmo}$ by
\begin{equation}\label{eq:hubblelaw}
    \textit{v}_\textrm{cosmo} = c \int_{0}^{\textit{z}_\textrm{cosmo}} \frac{\textrm{d}\textit{z}}{E(\textit{z})},
\end{equation}
where we have used the dimensionless Hubble parameter $E(\textit{z}) = H(\textit{z})/H_0$ \citep{Harrison93,Peebles93}.
The parameter $\textit{z}_\textrm{cosmo}$ is not directly measured, but can be determined from the observed redshift corrected to the cosmic microwave background (CMB) rest-frame $\textit{z}_\textrm{CMB}$ and the peculiar redshift approximated as $\textit{z}_\textrm{pec}\approx\textit{v}_\textrm{pec}/c$ 
(an approximation to the special relativistic Doppler shift)
such that:
\begin{equation}\label{eq:zcosmo_zpec}
    1+\textit{z}_\textrm{cosmo} = \frac{1+\textit{z}_\textrm{CMB}}{1+\textit{z}_\textrm{pec}}.
\end{equation}
The approximation $\textit{z}_\textrm{pec}\approx\textit{v}_\textrm{pec}/c$ is justified because PVs only reach $\sim 600$ km~s$^{-1}$.

\begin{figure*}[hbtp]
    \centering
    \includegraphics[width=480pt]{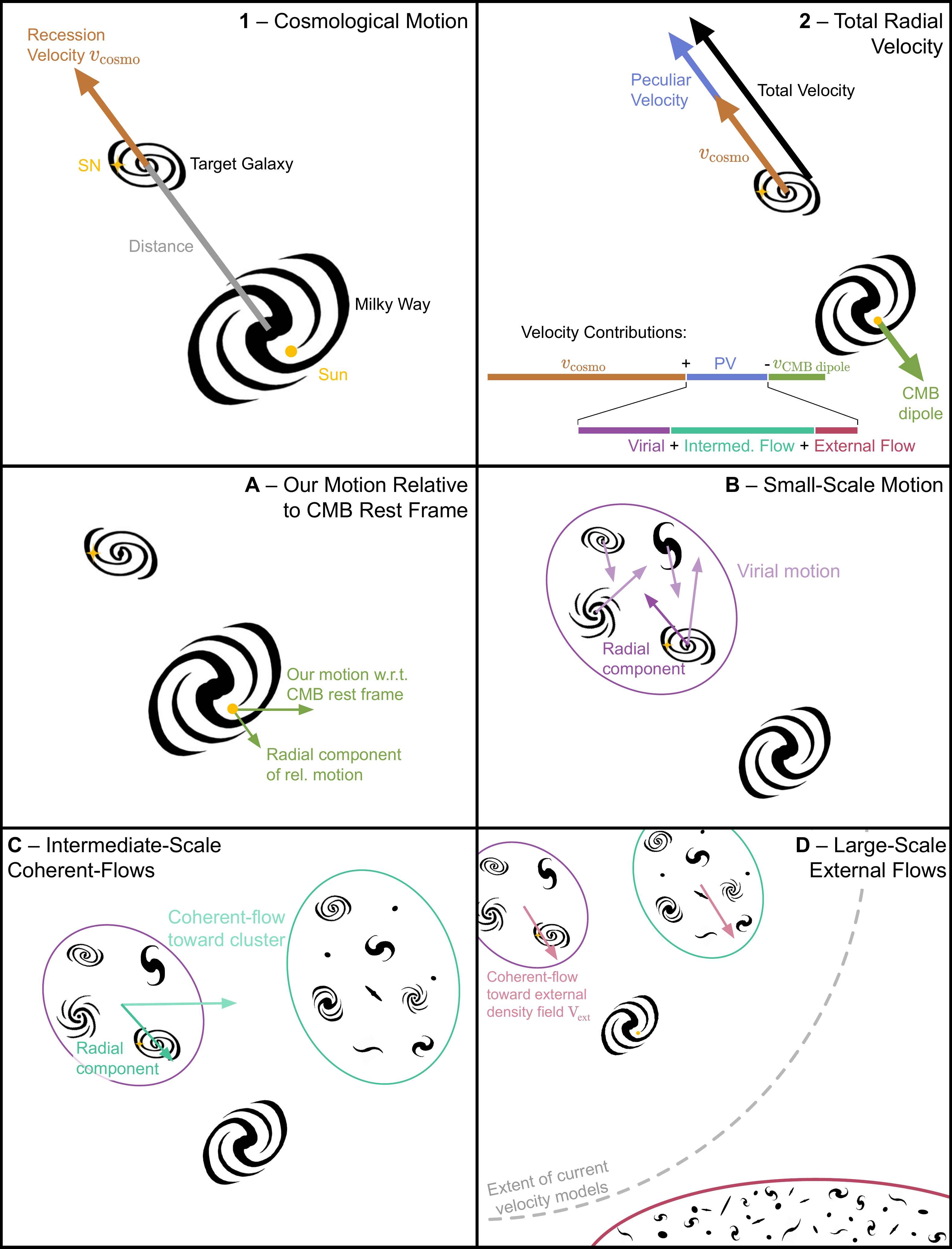}
    \caption{Explanation of the difference between the apparent cosmological motion of galaxies due to the expansion of the universe and its relation to distance (\textbf{Panel 1}) versus the observed motion of galaxies and peculiar motions (\textbf{Panel 2}).  The corrections for motions are broken into four components in \textbf{Panels A--D} and include: (\textbf{A}) relative CMB motion due to the motion of our galaxy, (\textbf{B}) small-scale virial motions of galaxies in a group, (\textbf{C}) large-scale coherent motions of halos of galaxies moving toward each other, and (\textbf{D}) large-scale external-field motions of all galaxies moving toward a place outside the local volume of redshift surveys. All of these motions (the final three being peculiar motions) contribute a radial component to the observed motion.}
    \label{fig:graphic}
\end{figure*}

PVs can arise from large-scale effects due to the coherent-flow (from here on designated as CF) motion of halos such as inflow into clusters or superclusters, and smaller-scale effects due to virial motion within galaxy clusters. These effects overlap somewhat in scale, but larger scale refers approximately to tens of $h^{-1}$ Mpc, and small scale to $<10$ $h^{-1}$ Mpc (where $h$ is defined as $H_0/(100\ \textrm{km}\ \textrm{s}^{-1}\ \textrm{Mpc}^{-1})$).
We provide a descriptive graphic in Fig.~\ref{fig:graphic} that explains the different components of observed galactic motion.
For convenience, many analyses distinguish between motions that occur due to the masses within the volume/field of the survey being analyzed and those due to masses beyond the extent of the survey. In total, we therefore consider four corrections applied in order to obtain better cosmological redshifts (the final three being contributions to PVs): our motion relative to the CMB rest frame, small-scale virial motion, larger-scale coherent motions, and large-scale external-field motions. These corrections are applied using

\begin{equation}\label{eq:zcosmo}
    z_\textrm{cosmo} = \bigg(\frac{(1+z_\textrm{observed})}{(1+z_\textrm{ecliptic})(1+z_\textrm{CMB dipole})(1+z_\textrm{pec})}\bigg) - 1
\end{equation}
\begin{equation}\label{eq:zpec}
    z_\textrm{pec} = \bigg((1+z_\textrm{virial})(1+z_\textrm{coh.})(1+z_\textrm{ext. coh.})\bigg) - 1,
\end{equation}
which can be approximated as:
\begin{equation}\label{eq:zcosmo_simple}
    z_\textrm{cosmo} = z_\textrm{observed} - z_\textrm{ecliptic} - z_\textrm{CMB dipole} - z_\textrm{pec} 
\end{equation}
\begin{equation}\label{eq:zpec_simple}
    z_\textrm{pec} = z_\textrm{virial} + z_\textrm{coh.} + z_\textrm{ext. coh.}.
\end{equation}

We do not use the approximated equations in this paper, but they are presented here to aid the reader.
The observed redshift $z_\textrm{observed}$ is first corrected to the Sun's rest frame by accounting for $z_\textrm{ecliptic}$ to retrieve a heliocentric redshift $z_\textrm{hel}$. Next, a correction to our motion relative to the CMB dipole $z_\textrm{CMB dipole}$ is accounted for to convert $z_\textrm{hel}$ to $z_\textrm{CMB}$. As shown in Eq.~\ref{eq:zcosmo_zpec}, we can obtain the cosmological redshift $z_\textrm{cosmo}$ by correcting for peculiar motions. In this paper, those peculiar motions include virial velocities ($v_\textrm{virial}$), which are motions within a group relative to the group's center of mass (we define $z_\textrm{group}$ as $z_\textrm{CMB}$ with $v_\textrm{virial}$ accounted for); CF velocities ($v_\textrm{coh.}$), which are motions of groups toward large-scale attractors (e.g.,~Virgo); and external CF velocities ($v_\textrm{ext. coh.}$), which are relative motions of groups toward ultra-large-scale attractors. We approximate peculiar redshifts with $\textit{v}_\textrm{pec} \approx \textit{z}_\textrm{pec} \times c$. A pictorial representation of these equations is given in Fig.~\ref{fig:graphic} (with $z_\textrm{ecliptic}$ not shown).

PVs ($\sim 300\ \textrm{km~s}^{-1}$) can contribute a significant fraction, up to $10\%$, of the overall apparent recession velocities at $z \sim 0.01$. Furthermore, PVs can be correlated across the sky and therefore may contribute systematic biases in $H_0$ or $w$ \citep{Neill07,Conley11}. At $z>0.1$, the statistical uncertainty in PVs is  $<1\%$ which is small compared to the relative distance-measurement uncertainty of $>6\%$.
It has been widely discussed in the literature that biased redshifts at low $z$ can propagate to significant uncertainties in the measurements of these cosmological parameters. \citet{Wojtak15} report that a redshift bias of $10^{-4}$ can bias $w$ by as much as 0.05, and similarly \citet{CalcinoDavis17} and \citet{Davis19} posit that a redshift bias of $5 \times 10^{-4}$ at low redshift can result in a bias of 1 km~s$^{-1}$ Mpc$^{-1}$ in $H_0$. 

Most recent cosmological analyses with SNe Ia attempt to correct for PVs using external catalogs
because significant non-zero measurements of the scale parameter for PV reconstructions, $\beta$ (the ratio of the growth rate of structure to galaxy bias $f(\Omega_m)/b$), imply these catalogs carry PV information that should improve scatter on the Hubble diagram \citep{Kaiser91,Hudson94,NusserDavis94}.
However, the accuracy and precision of PV catalogs are uncertain.
SNe have also been used as PV tracers and have also constrained the value of $\beta$, thus encouraging the implementation of PVs \citep{Riess97,PikeHudson05,Carrick15,Boruah20a}.
Systematic uncertainties in the PV measurements have typically been included in analyses measuring $w$ with SNe.
\cite{Scolnic18}, 
by varying $\beta$ by $10\%$, account for one possible systematic uncertainty in CF corrections 
and find that this causes an uncertainty in $w$ of 0.003. 
\cite{Scolnic18} showed the impact of omitting PV corrections shifts $w$ by $+0.009$. Following \citet{Scolnic18}, \citet{Brout19} find 
that the uncertainties in CF-corrected redshifts propagate to a systematic uncertainty in $w$ of 0.007. 
They add an additional systematic uncertainty due to a possible bias in redshifts of $4\times10^{-5}$ which they find contributes an uncertainty in $w$ of 0.006. To assess the impact of PV corrections on measurements of $H_0$, \cite{Riess16} explore the change in $H_0$ when the minimal redshift cut used for the sample is $z=0.01$ versus $z=0.023$. \cite{Riess16} use $z=0.023$ as the nominal cut to reduce sensitivity to PVs and find that this results in a difference in $H_0$ of 0.3 km~s$^{-1}$ Mpc$^{-1}$ compared to when using a minimum redshift of $z=0.01$.

The uncertainties in cosmological redshifts at low $z$ come primarily from two sources: measurement uncertainty and PV uncertainty.  For measurement uncertainties, these are typically on the order of $\sigma_z=10^{-4}$ as the redshifts used are mostly spectroscopic redshifts of the host galaxies.
With the exception of a minority of the sample ($\sim10\%$), the measurement uncertainties are subdominant to the PV uncertainties.  
\citet{Scolnic18} derived an uncertainty in the PVs at low $z$ of 250 km~s$^{-1}$ ($\sigma_z\approx0.0008$) by comparing the scatter of distance modulus residuals at $z\sim0.01$ to $z\sim0.05$. A similar magnitude for $\sigma_z$ was found in \citet{Burns18} who solved for the PV uncertainty simultaneously in the global $H_0$ fit and measured uncertainties in the PVs from 200--300 km~s$^{-1}$. Other studies have obtained either approximately equal or lower PV uncertainties \citep{Willick97,Pesce20,Blakeslee21,HollingerHudson21}. Interestingly, \cite{Scolnic18} find an individual redshift uncertainty of 260 km~s$^{-1}$ without large-scale CF corrections versus 250 km~s$^{-1}$ with the corrections. This shows that, while past corrections were reducing scatter, they were not reducing much of~it.

PVs to account for CFs can generally be derived in two ways.  The first method is from a density field, such as  2M++ \citep{Lavaux11,Carrick15}.\footnote{2M++ relies on observations from the Two-Micron-All-Sky Extended Source Catalog \citep[2MASS-XSC;][]{Skrutskie06} and redshifts from the 2MASS Redshift Survey \citep[2MRS;][]{Huchra12,Macri19}, the 6dF Galaxy Survey \citep[6dF; ][]{Jones09}, and the Sloan Digital Sky Survey \citep[SDSS-DR7;][]{Abazajia09}.}
The PV fields depend on relations that describe how galaxies trace the total matter present. One must assume mass continuity and standard gravity in an expanding universe to relate PVs to the gravitational accelerations derived from the density field. 
\citet{Carrick15} also present an uncertainty due to the smoothing scale of their PV field of 150 km~s$^{-1}$. This is significantly smaller than the 250 km~s$^{-1}$ found in \citet{Scolnic18}.

The second method for obtaining CF corrections is from distance and redshift surveys that make direct measurements on the PVs themselves, like those done by \textit{Cosmicflows}-3 \citep[\textit{Cf3};][]{Tully16} and the 6dF Galaxy Survey \citep[6dF;][]{Springob14}.
The 6dF is used to construct the PV field by
measuring each galaxy's distance from a fundamental plane (FP) relation and assuming that offsets from the FP are due to PVs \citep{Springob14}. \textit{Cf3} combines their previous data sets \citep[derived mostly from luminosity-linewidth correlations, also known as the Tully-Fisher relation, and the FP relation;][]{Tully13} with 6dF. 
Furthermore, as shown in Fig.~\ref{fig:graphic}.D, most models of the PV field based on a density field attempt to determine a coherent motion of the galaxies inside the survey volume toward a place outside this volume (albeit \textit{Cf3} makes no distinction between an internal and external CF). Differences in the measurement of this external CF are discussed in our analysis.

While large CF corrections have received a significant amount of attention in the literature, corrections due to galaxy-group corrections have received less, specifically in the SN community. 
Galaxy groups are typically defined as associated galaxies from virial motion and relative distances, but the galaxy-group assignments themselves are not agreed upon in the literature.
Corrections for these motions are important because they remove the virial velocities of galaxies relative to the group center.
Group-averaged redshifts have been used for a fraction ($\sim10\%$) of the low redshifts in Pantheon based on a collection in the NASA/IPAC Extragalactic Database (NED) used by \citet{Carrick15}.
\citet{Qin18} also implement groups to compare the 2MASS Tully-Fisher Survey to the 6dF in order to average redshifts within groups and avoid using misclassified galaxies. \citet{Tully16} use groups assigned by \citet{Tully15} to obtain group-averaged distances and velocities, compare with other group catalogs, and improve upon previous large-scale flow studies. 
In these catalogs, no information is given on the probability of a galaxy being in a group --- a galaxy is either in a group or not.

While typically PVs are used in SN cosmological analyses to constrain cosmological parameters, here we test multiple sets of PVs as corrections to redshifts and observe their impact on SN cosmology.
Previous works have studied several PV catalogs and methods \citep{Boruah20b,Blakeslee21,Rahman21}, but we aim to cover a more diverse range of catalogs here. 
Importantly, the analysis done here uses the same sample that is used for both \citet[][SH0ES]{Riess21} and Brout et al.~2022 in prep.~and \citet[][hereafter Pantheon+]{Scolnic21}.
This paper is a companion paper to \citet{Carr21}, which does a full review of all redshift samples and also derives the 2M++ velocities used in this work.
In this paper, we study the efficacy of both group and CF corrections using a current compilation of low-$z$ SNe Ia. We explain our data sample and light curve fits in section~\ref{sec:Data}. In section~\ref{sec:Group_Analysis},
we review various methods of group assignment. We perform a similar exercise for various CF methods in section~\ref{sec:Pec_Vel_Analysis}. In section~\ref{sec:Results}, we explain which sets of redshift corrections reduce scatter on the Hubble diagram and determine the potential impact on $H_0$ and $w$ from these different corrections. Finally, discussion and conclusions are presented in sections~\ref{sec:Discussion} and~\ref{sec:Conclusions}.

\begin{figure}
    \centering
    \includegraphics[width=\columnwidth]{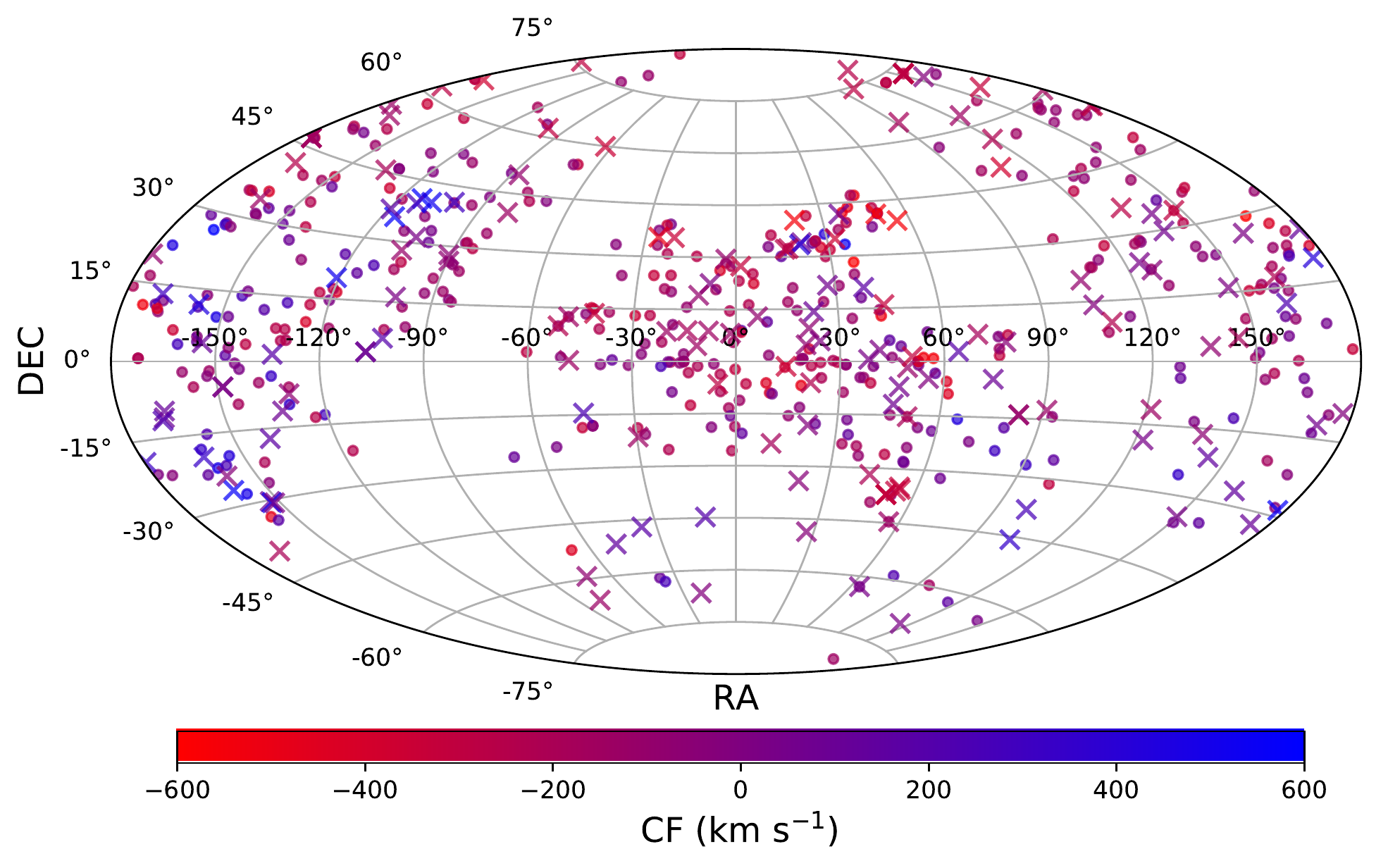}
    \caption{Distribution of the low-$z$ SNe ($z < 0.08$) in our sample across the sky. The points are color-coded by the CF corrections (in km~s$^{-1}$) obtained using the 2M++ density field. The $\times$'s correspond to those SNe we find to be in a group in any of the \citet{Tully15}, the \citet{Lim17}, the \citet{Lambert20}, or the \citet{Crook07} group catalogs.}
    \label{fig:lowz_vpec_colorbar}
\end{figure}

\section{Data}\label{sec:Data}
We use a large compilation of low-redshift SNe to analyze various PV treatments. We follow \citet{Kenworthy19} who augment the low-redshift sample from Pantheon \citep{Scolnic18} with an updated sample of Carnegie Supernova Project (CSP) SNe from \citet{Krisciunas17} as well as the Foundation sample \citep{Foley18,Jones19}. We also add in the Lick Observatory Supernova Search (LOSS) sample \citep{Stahl19} which contributes $\sim100$ SNe Ia at $z<0.08$.  The Pantheon sample has low-$z$ SNe from the Center for Astrophysics data sets  \citep[CfA1-4; ][]{Riess99,Jha06,Hicken09a,Hicken09b,Hicken12} as well as a small amount of Pan-STARRS1 \citep[PS1; ][]{Scolnic18}, Swift Optical Archive \citep[SWIFT; ][]{swift}, Complete Nearby ($z<0.02$) Sample of Type Ia Supernova Light Curves \citep[CNIa0.02; ][]{asassn}, SDSS \citep{Sako11}, and SNLS \citep{Betoule14} low-$z$ SNe. A full presentation of the light curves used in this analysis (including high-$z$ SNe) will be given in Pantheon+.
The positions of the SNe across the sky are presented in Fig.~\ref{fig:lowz_vpec_colorbar}. SNe are more often found toward the Northern Hemisphere due to the location of the telescopes that discovered them.

\subsection{Redshift Sources}
Our initial heliocentric redshifts for this compilation of SNe come largely from Pantheon and references therein, as well as from \citet{Krisciunas17}, \citet{Foley18}, and \citet{Stahl19} for CSP, Foundation, and LOSS respectively. At low $z$, redshifts in Pantheon have been primarily obtained from host galaxy spectra, and $<$~$6\%$ from the SN spectral features themselves \citep{Steinhardt20}. Here we use the redshifts that we present in a companion paper \citet{Carr21}, which examines and reassigns the host galaxies of all the low-$z$ SNe and query NED for the most recently acquired redshift of all the galaxies.

For converting heliocentric redshifts into redshifts in the CMB frame, we use the formula
\begin{multline}\label{eq:zhelzcmb}
    \textit{z}_{\textrm{CMB}} = (1+\textit{z}_{\textrm{hel}}) / \bigg[ 1-\frac{v_\textrm{0 CMB dipole}}{c}  \big(\sin(b) \sin(b_0)\\
    + \cos(b) \cos(b_0)
    \cos(l-l_0) \big) \bigg]-1,
\end{multline}
where $l$ is the longitude, $b$ is the latitude of the observed galaxy in galactic coordinates, the combination of $l_0$ and $b_0$ is the dipole location with $l_0$ =~264.021$^{\circ}$ and $b_0$ =~48.253$^{\circ}$, and $v_\textrm{0 CMB dipole}$ =~369.82 km~s$^{-1}$ is the velocity due to the Sun's motion with respect to the CMB rest frame \citep{Planck18}.

\subsection{Measuring Distance Modulus Values}
To measure distance modulus values for the SNe, we use \texttt{SNANA} \citep{Kessler09} to fit the SN light curves with the SALT2 model \citep{Guy10,Betoule14}. The fits return an overall normalization $x_0$ related to the apparent peak brightness $m_B$, a stretch factor $x_1$, and a color parameter $c$ for each SN. 

We apply the same light curve quality cuts as described in \citet{BroutScolnic21}. Criteria include $|x_1| < 3$ with an uncertainty $< 1$, $|c| < 0.3$ with an uncertainty $< 0.05$, uncertainty in peak magnitude date $< 2$ days, Milky Way (MW) extinction $< 0.2$, and light curve fit probability (from \texttt{SNANA}) $> 0.01$. After these quality cuts, in total, this low-$z$ ($z < 0.08$) sample consists of $\NSNeincoherents$ unique SNe. 

To convert the fitted parameters to a distance modulus $\mu$, we follow a modified version of the Tripp estimator \citep{Tripp98} as given in \cite{Scolnic18} where 
\begin{equation}
   \mu = m_B + {\alpha}x_1 - {\beta}c + \gamma - \Delta_B - \mathcal{M} .
\end{equation}
Here, $m_B$ represents the apparent peak brightness, $\alpha$ and $\beta$ are correlation coefficients relating $x_1$ and $c$, respectively, to luminosity, $\gamma$ is the correction for the mass-luminosity relation \citep[typically called the mass-step;][]{Kelly10,Lampeitl10,Sullivan10}, $\Delta_B$ is the bias correction based on the BBC method using simulations \citep[BBC method;][]{KesslerScolnic17}, and $\mathcal{M}$ is the absolute brightness of a SN with $c~=~0$ and $x_1~=~0$. Following \citet{Kenworthy19}, we fix $\alpha = 0.14$ and $\beta = 3.1$, and $\gamma$ =~0.06 mag such that a luminosity step of +0.03 mag is applied for mass $ >10^{10} M_{\odot}$ and $-0.03$ mag for mass $ <10^{10} M_{\odot}$. We take the assigned stellar masses from \citet{Jones18} for both Pantheon and Foundation and use \citet{Kenworthy19} for stellar masses for the CSP sample. The derivations of other masses are explained in Pantheon+. The simulations used for the bias correction are described in \cite{Scolnic18}, \cite{Jones18}, and \cite{Kessler19}, and the color and stretch parameters are described in \citet{Popovic20}.

Following \citet{Scolnic18} we use 
\begin{equation}\label{eq:distance_error}
    \sigma_{\mu}^2 = \sigma_\textrm{N}^2 + \sigma_{\mu-z}^2 + \sigma_\textrm{int}^2,
\end{equation}
to calculate the total distance error ($\sigma_{\mu}$) for the SNe in our sample by combining in quadrature: $\sigma_\textrm{N}$ the measurement uncertainty of the SN distances based on $m_B$, $x_1$, and $c$; $\sigma_{\mu-z}$ the uncertainty from the PV uncertainty and redshift measurement uncertainty; and $\sigma_\textrm{int}$ the intrinsic scatter which is further discussed in section~\ref{subsec:improv_hubble_res}.

\begin{deluxetable*}{ccc}
\tabletypesize{\footnotesize}
\tablenum{1}
\tablecaption{Group Corrections (N = \NSNeingroups)} \label{tab:group}
\tablewidth{0pt}
\tablehead{\colhead{Group Variant} & \colhead{Explanation} & \colhead{N (matched)}}
\startdata
Gal  &  Groups not implemented & \NSNeingroups\\
T15  &  Velocity and distance association \citep{Tully15} & \NSNeintully\\
Lim17  & Halo-based group finder \citep{Lim17} & \NSNeinlim\\
Lam20 & Friends of Friends algorithm \citep{Lambert20} & \NSNeinlam \\
C07  & Maximizing groups with three or more galaxies \citep{Crook07} & \NSNeincrook\\
\enddata
\tablecomments{Group acronyms are written with a short description and number of SNe matched.}
\end{deluxetable*}

\section{Group Analysis}\label{sec:Group_Analysis}

\subsection{Group Assignments}
We use groups identified by \citet{Tully15}, \citet{Lim17}, \citet{Lambert20}, and \citet{Crook07} each individually in an attempt to calculate $z_\textrm{group}$, which will improve our estimate of $z_\textrm{cosmo}$ by removing the virial motion’s contribution to the observed redshift (Fig.~\ref{fig:graphic}.B).
A summary table is given in Table~\ref{tab:group}. 
\citet[][hereafter T15]{Tully15} defines galaxy `nests' or galaxy groups by associating galaxies in velocity and distance space and iteratively improving the group membership by redefining the center of the group and ensuring no overlaps occur between groups (see section~4 of T15 for details).
The redshifts for this group catalog come from the 2MASS Redshift Survey (2MRS; \citealp{Huchra12}).

\citet[][hereafter Lim17]{Lim17} define their groups using a halo-based group-finder method, which identifies groups based on dark matter halo properties, such as mass and velocity dispersion. Additionally, Lim17 demonstrate their rigorousness by testing their group finding method on simulations and observing about a 94\% success rate among mock samples. Redshifts for this group catalog come primarily from 2MRS but some also come from 6dF \citep{Jones09}, SDSS \citep{Albareti17}, and the Two Degree Field Galaxy Redshift Survey \citep[2dFGRS; ][]{Colless01}. 

\citet[][hereafter Lam20]{Lambert20} also use 2MRS to construct a group catalog. They implement a modified Friends of Friends algorithm \citep[FoF; ][]{HuchraGeller82} based on graph theory with which they attempt to improve upon previous group catalogs by both keeping large clusters from being mistakenly broken down into smaller groups and avoiding the misclassification of large clusters from small groups at the same time.

\citet[][hereafter C07]{Crook07} assign groups to the 2MASS-XSC using an earlier release of 2MRS \citep{Huchra05a,Huchra05b} by comparing galaxy distances and velocities through a FoF algorithm. They define two galaxies to be in a group if the galaxies are within a linking distance (dependent on density) of their average redshift and their difference in redshift is within the defined linking velocity. We take the low-density catalog from C07 who define the fiducial linking distance as 1.63 Mpc and fix the linking velocity at 399 km~s$^{-1}$ thereby maximizing the number of groups with three or more members.

\subsection{Group Identification Methods}

\begin{figure}
    \centering
    \includegraphics[width=\columnwidth]{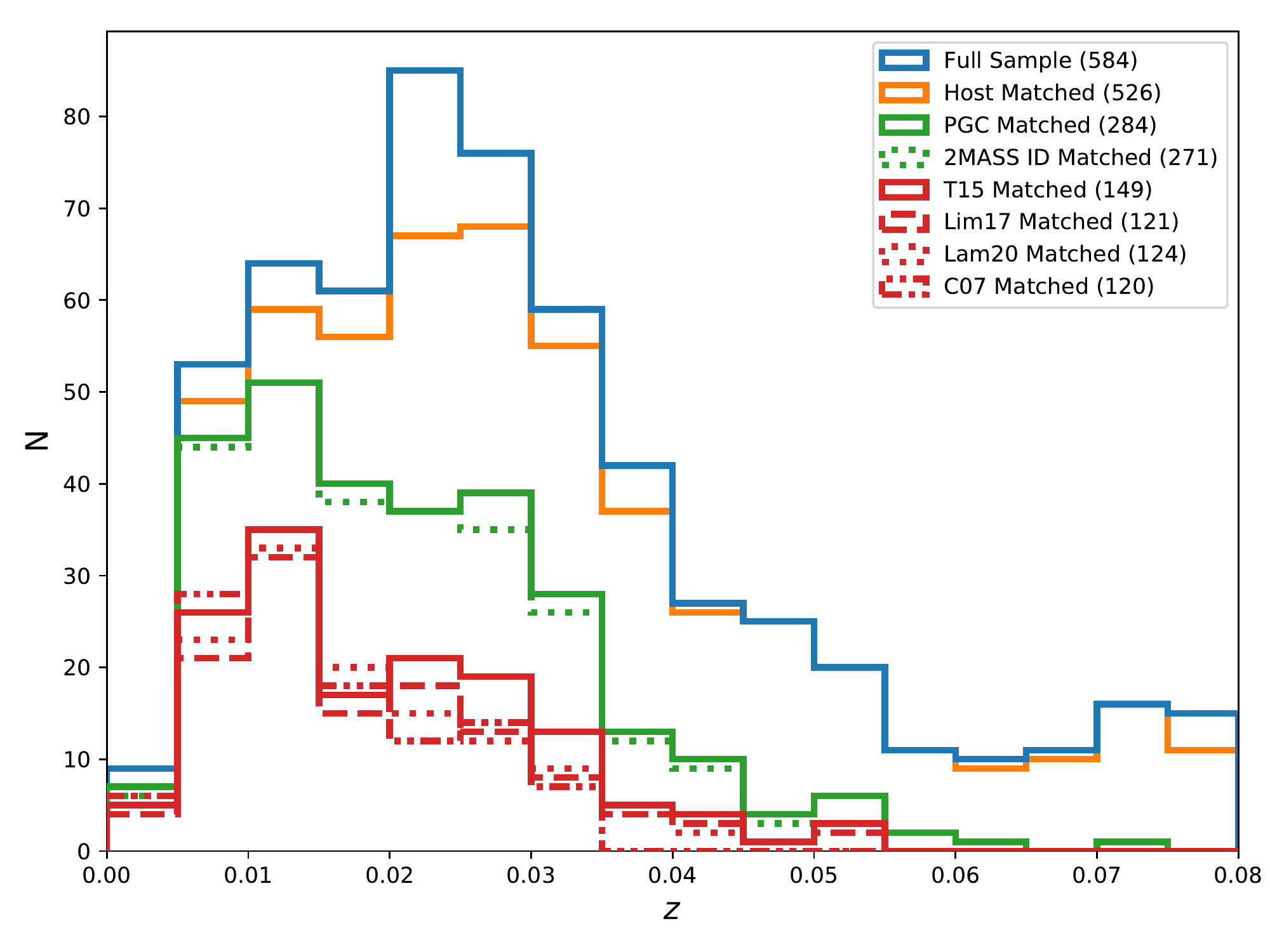}
    \caption{Histogram of the number of galaxies versus redshift, after successive assignments toward group catalogs. The full sample is shown in blue. Those that have host galaxy identifications are shown in orange. Hosts that appear in the PGC or 2MASS catalogs are in green (with 2MASS dotted). These can then be matched to the existing group catalogs in red (T15, Lim17, Lam20, and C07).}
    \label{fig:z_cuts}
\end{figure}

\begin{figure*}[!htb]
    \centering
    \includegraphics[width=\textwidth]{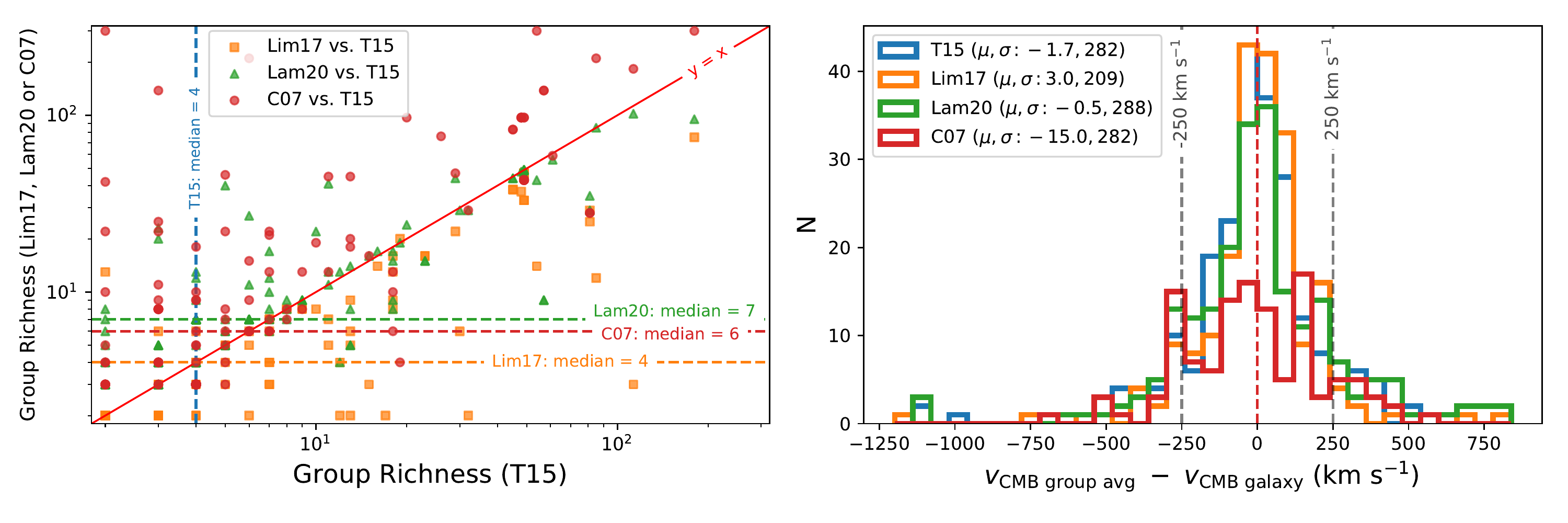}
    \caption{Comparing \citet{Tully15}, \citet{Lim17}, \citet{Lambert20}, and \citet{Crook07} group assignments for the SNe found in groups. \textbf{Left}: Corresponding group richnesses for the galaxies matched in our sample on a logarithmic scale.
    \textbf{Right}: Histogram of the unweighted average redshifts of a group subtracted by the host galaxy's redshift for each group catalog. Median differences ($\mu$) and standard deviations ($\sigma$) are given in km~s$^{-1}$ in the legend.}
    
    \label{fig:comparing_lim_tully}
\end{figure*}

The process for identifying galaxy groups is depicted in Fig.~\ref{fig:z_cuts}.
In order to use these grouping catalogs, we match the SNe (N =~\NSNeincoherents) to their host galaxies as determined by \citet{Carr21}. 
Any galaxy labeled to be in a group in T15 or Lim17 has an identification in the Principal Galaxies Catalog (PGC), so we match our host galaxies to entries in the PGC using the Lyon Extragalactic Database\footnote{\url{http://leda.univ-lyon1.fr/}}. We then use the Extragalactic Distance Database\footnote{\url{http://edd.ifa.hawaii.edu/dfirst.php}}, which has taken a number of different catalogs and matched according to PGC numbers, to compare to T15
and Lim17. 
We successfully find $\NSNeintully$ SNe that are in groups in T15 and $\NSNeinlim$ in Lim17. The corresponding number of matched galaxies for Lam20 is $\NSNeinlam$ (matching by 2MASS IDs), where groups are predefined as having three or more galaxies. We match $\NSNeincrook$ SNe to galaxies in C07 by minimizing coordinate distance (defining an appropriate host galaxy if the coordinate separation is $< 0.01^{\circ}$).

We plot any SN host galaxy deemed to be in a group in T15, Lim17, Lam20, or C07 in Fig.~\ref{fig:lowz_vpec_colorbar} alongside galaxies that do not have any identified group. 
We determine that up to 30\% (N =~$\NSNeingroups$) of our low-$z$ galaxies can be assigned to groups and note a selection bias toward the lowest redshifts due to the completeness of the catalogs used.
We find that the group assignments are not biased toward a specific region in the sky. Separating the sky into four quadrants, we find that each quadrant has $28.4\% \pm 3.2\%$ of SNe found in groups.

The differences in group richness (the number of galaxies per group) between T15 and Lim17, Lam20, or C07 for the SNe in a group in any of the four catalogs is illustrated in the left panel of Fig.~\ref{fig:comparing_lim_tully}. The median group richness is 4 for T15, 4 for Lim17, 7 for Lam20, and 6 for C07. For this sample of galaxies, 39\% of groups are richer in T15 as compared to Lim17's (58\% are defined as equally rich). That number drops to 14\% when comparing T15 group richness to Lam20 group richness (24\% have equal richness). Finally, 51\% of T15 groups are richer than C07's (with 9\% of galaxies defined to be in groups of the same richness).

\subsection{Determining Redshifts of Groups}\label{subsec:group_utilize}

Given these group assignments, there are a number of different options for determining the group redshift: using a simple average (Group Avg), taking the brightest galaxy's redshift (Brightest), and taking a weighted mean (Group Mass Weight).

The first method for obtaining group redshifts is an unweighted average of all galaxies in a given group. 
We compare group-averaged redshifts and galaxy redshifts in the right panel of Fig.~\ref{fig:comparing_lim_tully} for \citet{Tully15}, \citet{Lim17}, \citet{Lambert20}, and \citet{Crook07}.
We show the median residuals in velocity between the galaxy redshifts from each catalog and the group-averaged redshifts
are on the order of 1--5 km~s$^{-1}$ (with the exception of C07) with standard deviations ranging from 200--300 km~s$^{-1}$.
We also calculate the redshift differences between the group-averaged redshifts of Lim17, Lam20, and C07 as compared to T15. Median differences between the various sets of group-averaged redshifts in velocities are $-9.1$, $-28.2$, and $31.0$ km~s$^{-1}$ respectively.

The second method we test for obtaining a group redshift is taking the redshift of the brightest galaxy in a group.
The third and final method we test is averaging according to mass.
We employ a pseudo-mass-weighted averaging method for the T15 group assignments. We weight according to the logarithmic intrinsic \textit{K}-band luminosity at the group distance (defined by the unweighted group velocity in the CMB frame divided by 100 Mpc). 
However, for Lim17, we weight according to the logarithmic stellar mass obtained by \citet{Lim17} from the relation between the stellar mass and \textit{K}$_s$-band luminosity.
For both T15 and Lim17, we obtain mass-weighted redshifts that are almost identical to the unweighted averaging.

For our group analysis, we use all three group redshift methods for both T15 and Lim17. For Lam20 and C07 we only use the simple average since we had no luminosity information for those catalogs. In total, we have eight group redshift variants.

\section{Coherent-Flow Analysis}\label{sec:Pec_Vel_Analysis}

\subsection{Explanation of CF Variants}\label{subsec:CF_variants}
To understand the impact of large-scale PV corrections, we obtain three sets of coherent-flow (CF) corrections for use in our analysis (Fig.~\ref{fig:graphic}.C, \ref{fig:graphic}.D).
The first set of corrections is based on 2M++, which we obtain using a velocity field reconstruction derived from the data from \citet{Lavaux11}. In order to extract the CF corrections from 2M++, we follow the general methodology in \citet{Carrick15}\footnote{\url{cosmicflows.iap.fr}} (hereafter C15) who create a predicted PV field as a function of position in real space, and \citet{Carr21} use that field and convert it to redshift space in two steps. They convert each grid point from a real-space position to a redshift-space position using the predicted CF correction at that point, and then they interpolate the resulting redshift-space grid using inverse distance weighting. By converting to a redshift-spaced PV field, the CF correction can be predicted at a given redshift as opposed to a presumed location, and biases are avoided. 
We analyze a number of variations that use 2M++ data:
\begin{itemize}
\item{2M++ C15: our nominal method, where the external CF model is determined in \citet{Carrick15} (CF corrections included in Fig.~\ref{fig:lowz_vpec_colorbar}).
The applied values of $\beta$ and $\textbf{\textit{v}}_\textrm{ext. coh.}$ are $\beta = 0.431$ and $\textbf{\textit{v}}_\textrm{ext. coh.}$~=~89$\textbf{\textit{\^{i}}}$ -~131$\textbf{\textit{\^{j}}}$ + 17$\textbf{\textit{\^{k}}}$ in km~s$^{-1}$ and galactic coordinates.}

\item{2M++$_\textrm{ne}$: similar to 2M++ C15, except `no external' CF field is applied \citep[i.e.,~$\textbf{\textit{v}}_\textrm{ext. coh.}$ = 0; ][]{Lavaux11}.}

\item{2M++$_\textrm{ilos}$: where CF corrections are calculated by integrating over all possible distances along the line of sight. We define a probabilistic model for the underlying cosmological redshift given the 2M++ reconstruction. The redshift-space density reconstruction gives a mapping for the predicted PV as a function of the cosmological redshift
$\textit{v}_\textrm{pec} ( z_\text{cosmo})$. Assuming a nonlinear velocity dispersion of $\sigma_v= 250$ km s$^{-1}$, we assume that the PV will be normally distributed about the predicted PV. We can then construct a likelihood function for the cosmological redshift, given the observed redshift, as
\begin{multline}\label{eq:darcy_pv}
    P(z_\textrm{CMB}|z_\textrm{cosmo}) = 1/\sqrt{2 \pi \sigma_v^2 } \cdot \exp (\\ - \left( (1+ \textit{v}_\textrm{pec} ( z_\text{cosmo}) /c  ) - \frac{1+z_\textrm{CMB}}{1+z_\textrm{cosmo}} \right)^2
    / (2 \sigma_v^2) ).
\end{multline}

We use a prior $P(z_\textrm{cosmo}) \propto (1 + \delta(z_\textrm{cosmo}) ) \frac{d \chi(z)}{dz}|_{z={z_\textrm{cosmo}}}  $ to account for the inhomogeneous Malmquist bias \citep{PikeHudson05}, where $\delta$ is the matter density contrast in the redshift-space reconstruction and $\chi(z)$ is the comoving distance as a function of redshift. We integrate over possible values of $z_\textrm{cosmo}$ by using the MCMC code \texttt{Stan} \citep{Betancourt17} to determine the posterior mean (given the observed redshift) of the cosmological redshift for each SN, which we use as a variant redshift vector.  In Appendix~\ref{sec:variablevariance}, we discuss use of the additional information available in the posterior distribution. This variant includes the external CF from 2M++ C15.} 

\item{2M++ [Gr]: where we calculate the 2M++ C15 CF correction at the group location (where available) rather than the individual galaxy location.}

\item{2M++/SDSS: similar to 2M++ C15, but applies $\beta$ and $v_\textrm{ext. coh.}$ values determined in \citet{Said20} from SDSS and 2M++ data. These are $\beta = 0.314$ and $\textbf{\textit{v}}_\textrm{ext. coh.}$ = 98$\textbf{\textit{\^{i}}}$ -~148$\textbf{\textit{\^{j}}}$ + 12$\textbf{\textit{\^{k}}}$. \citet{Said20} explain that they compare inferred PVs obtained from the FP relation using SDSS, 6dF, or a combination of the two to PVs inferred from the 2M++ density field. Using this comparison \citet{Said20} simultaneously fit for $\beta$, $v_\textrm{ext. coh.}$, and FP relation coefficients.}

\item{2M++/SDSS$_\textrm{ilos}$: which is the same as 2M++/SDSS but integrating over all possible distances along the line of sight as in 2M++$_\textrm{ilos}$.}

\item{2M++/SDSS/6dF: similar to 2M++/SDSS, but using the $\beta$ and $v_\textrm{ext. coh.}$ determined with SDSS, 2M++, and 6dF data, as done in \citet{Said20}. These are $\beta = 0.341$,  $\textbf{\textit{v}}_\textrm{ext. coh.}$ = 94$\textbf{\textit{\^{i}}}$ -~138$\textbf{\textit{\^{j}}}$ + 4$\textbf{\textit{\^{k}}}$.} 

\end{itemize}

The second set of CF corrections is from \textit{Cf3} using the methodology of forward modeling the data set as described in \citet{Graziani19}.
This set of PVs was computed with T15 group information where available.
For the third set, we compare to a simpler model from \citet{Mould00} who focus on CFs derived almost exclusively from the Virgo cluster, the Great Attractor, and the Shapley supercluster (hereafter VGAS).
Last, we include a recent set from \citet{LilowNusser21} who obtain a constrained realization of the PV field from 2MRS by utilizing both a variance-minimizing Wiener filter and random residual field realizations\footnote{\url{https://github.com/rlilow/CORAS}}. We call this set of CF corrections 2MRS. 
We also analyze a 2MRS integrated line-of-sight variant named 2MRS$_\textrm{ilos}$.
Unless otherwise specified, we use host galaxy positions and redshifts to determine the corresponding CF corrections from each set.

\begin{figure}[!htb]
    \centering
    \includegraphics[width=\columnwidth]{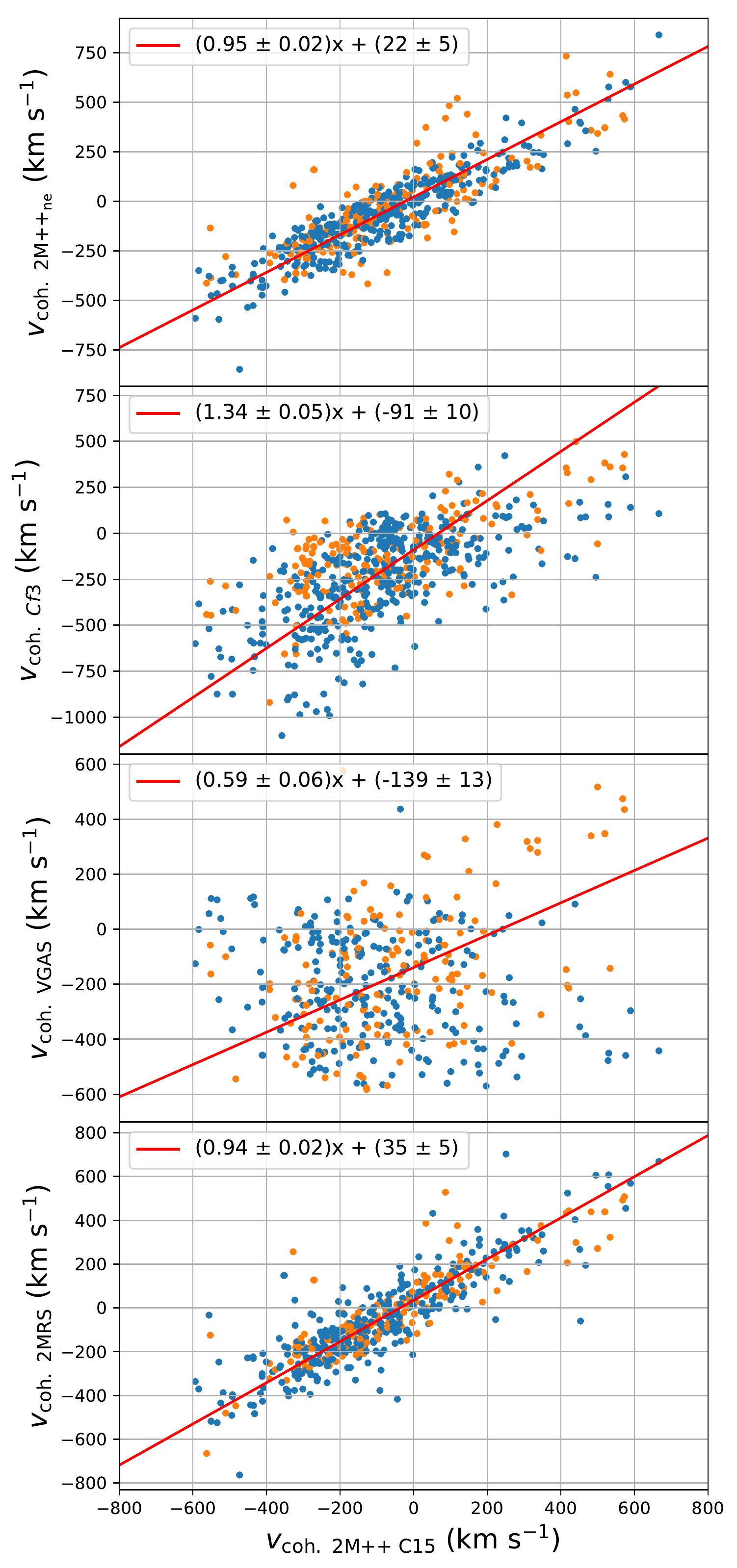}
    \caption{Comparison between $v_\textrm{coh.}$ from the primary sets of CF corrections. Results from an orthogonal distance regression fit are shown assuming no uncertainties in the PVs. SNe with $z < 0.02$ are in orange. All others are in blue. \textbf{Upper}: Comparing 2M++$_\textrm{ne}$ to 2M++ C15. \textbf{Upper Middle}: Comparing \textit{Cf3} to 2M++ C15. \textbf{Lower Middle}: Comparing VGAS to 2M++ C15. \textbf{Lower}: Comparing 2MRS to 2M++~C15.}
    \label{fig:cf3_vpec_residual}
\end{figure}

\subsection{CF Catalog Comparison}
We compare five variant sets of CF corrections and fit a line to each to show differences in methodology and assumptions in Fig.~\ref{fig:cf3_vpec_residual}. 
We compare two of the variant CF correction sets from 2M++ (with and without an external CF) and find a correlation, using an orthogonal distance regression, with a slope of $0.95 \pm 0.02$, as shown in the upper panel of Fig.~\ref{fig:cf3_vpec_residual}.
In the upper-middle panel of Fig.~\ref{fig:cf3_vpec_residual}, we find a correlation between the CF corrections reported by 2M++ C15 and the extracted CF corrections from \textit{Cf3} observing a slope of $1.34 \pm 0.05$ and an offset of $91 \pm 10$ km~s$^{-1}$. 
Some potential explanations for this offset between 2M++ and \textit{Cf3} CFs, which is worse at higher redshifts, are improper Malmquist Bias corrections at the large distances of the \textit{Cf3} data set, differences between methodologies based on redshift surveys and direct PV measurements, and asymmetry in the SN sample, which is dominated by the Northern Hemisphere.  
This will be under further investigation in the fourth generation of the \textit{Cosmicflows} catalog, which contains about 45,000 individual galaxy distances.
In the lower-middle panel of Fig.~\ref{fig:cf3_vpec_residual}, when we compare the set from 2M++ C15 to \citet{Mould00}, we see little correlation between the $v_\textrm{coh.}$'s ($0.59 \pm 0.06$ and an offset of $139 \pm 13$ km~s$^{-1}$). 
This is likely because \citet{Mould00} only consider the three largest superclusters and therefore do not capture most of the CFs across the sky.
Finally, in the lower panel we compare 2MRS with 2M++ C15 and observe a slope of $0.94 \pm 0.02$ and an offset of $35 \pm 5$. We further discuss the implications of some of these velocity offsets in sections~\ref{sec:Results} and~\ref{sec:Discussion}.

\section{Results}\label{sec:Results}
\subsection{Improvement in Hubble Residuals}\label{subsec:improv_hubble_res}

To determine the efficacy of the various PV corrections, we measure the improvement in the $X^2$ and in the dispersion of the Hubble residuals about a fiducial cosmology.  
This is based on the assumption that removing PVs should reduce the dispersion in the Hubble diagram.
We take the distance moduli ($\mu$) and the uncertainties in the distance moduli ($\sigma_{\mu}$) from Eq.~(\ref{eq:distance_error}) 
to compute Hubble residuals and an $X^2$ relative to a fiducial cosmology following,

\begin{equation}\label{eq:mu_dif}
    \Delta_{\mu,i} = \mu_{\textrm{obs,}i}-\mu_{\textrm{cosmo}}(z_i)
\end{equation}
\begin{equation}\label{eq:chi2}
    X^2 = \sum_i{\frac{\Delta_{\mu,i}^2}{\sigma_{\mu,i}^2}},
\end{equation}
where the index $i$ runs over all SNe in the sample, $\mu_\textrm{obs}$ is the observed distance modulus, and $\mu_\textrm{cosmo}(\textit{z})$ is the predicted distance given a redshift $z$ and the best-fit $\Lambda$CDM parameters from the Pantheon+ analysis to calculate the distance modulus residual ($\Delta_{\mu}$).
Although this statistic is often referred to as a $\chi^2$ in the literature, we call it $X^2$ as we make no claims about its distribution under a null hypothesis. This statistic does not take account of correlations present in the data from effects such as calibration, bias correction models, or MW extinction (see Brout et al.~2022 in prep.~for a full analysis). However, as we are principally interested in the diagonal scatter caused by PVs, these effects have little impact on our final conclusions.

While we do not change the best-fit cosmology that we use to measure $X^2$ when revising redshifts based on different PV analyses, we note that allowing the best-fit cosmology to vary does not alter any of the trends found.
We incorporate a $\sigma_\textrm{int}$ (introduced in Eq.~\ref{eq:distance_error}) of $\groupsigint$ mag in both our group and CF analyses so that the reduced $X^2$ of our distance modulus residuals is close to 1.
This sets a consistent $X^2$ floor for our $X^2$ comparisons.
Following \citet{Scolnic18} we assume a PV uncertainty of 250 km~s$^{-1}$. This assumption is further discussed in section~\ref{sec:coherent-flow_scatter} and Appendix~\ref{sec:variablevariance}.

We apply an outlier cut to the distance modulus residuals at $\chaevenautscrit \sigma$ relative to the best-fit cosmological model; this is more relaxed than past analyses which use Chauvenet's criterion at $3.5\sigma$ \citep{Scolnic18} because we found the Hubble residuals can change significantly when the redshift is changed.
For this analysis, we include only SNe that pass this 5$\sigma$ outlier cut for our baseline set of redshifts
where $z_\textrm{hel}$ is taken from the galaxy and no CF correction is applied ($z_\textrm{hel}$ =~Gal and CF~=~None).
Applying the typical cut at $3.5\sigma$ rather than $5.0\sigma$ did not significantly change our results.

\begin{figure*}[!htb]
    \centering
    \includegraphics[width=\columnwidth]{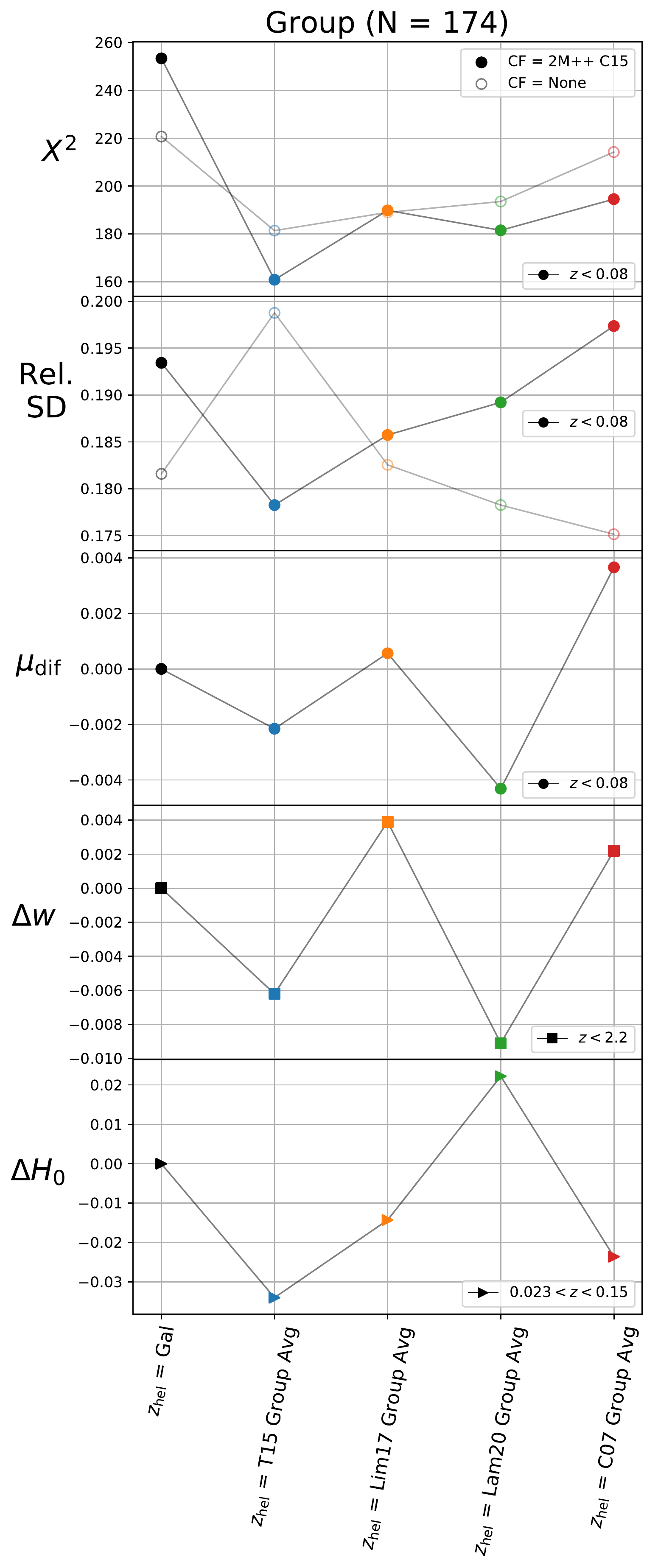}
    \includegraphics[width=\columnwidth]{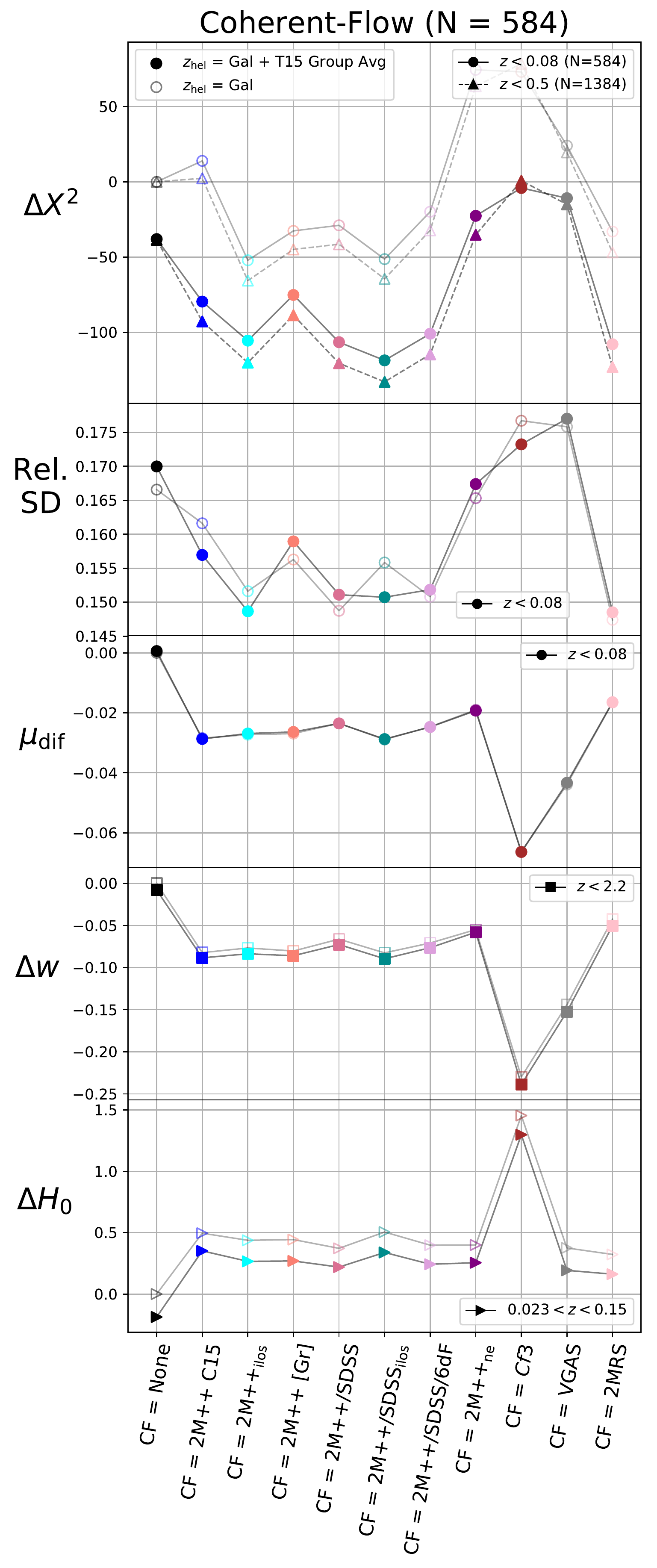}
    \caption{Results from our fitting analysis. 
    Descriptions of all variants are provided in Table~\ref{tab:group} and section~\ref{subsec:CF_variants}. Lines are used for visualization only. \textbf{Left}: Group $X^2$, relative standard deviation of the Hubble residual (Rel.~SD), difference in weighted mean distance modulus residuals ($\mu_\textrm{dif}$), $\Delta w$, and $\Delta H_0$ for \citet{Tully15}, \citet{Lim17}, \citet{Lambert20}, and \citet{Crook07} group-averaged redshift variants as compared with the galaxy redshift variant. We compare $X^2$ values with (solid points) and without (open points) an additional CF correction. In the first, second, and third panels we present values calculated by using 
    only SNe updated as described in our group analysis (N =~$\NSNeingroups$). In all panels, only SNe found in groups (N~=~$\NSNeingroups$) vary between points on a line with some panels showing values calculated with samples covering differing redshift ranges.
    The fourth and fifth panels are calculated with $z < 2.2$, and $0.023 < z < 0.15$ respectively.
    \textbf{Right}: Comparing $\Delta X^2$, Rel.~SD, $\mu_\textrm{dif}$, $\Delta w$, and $\Delta H_0$ for various CF corrections applied to the individual galaxy redshifts or group-averaged heliocentric redshifts from T15, where available. Additionally plotted with open points are redshift sets with solely individual galaxy redshifts. In the first panel, the improvements in $X^2$ for two upper limits of redshift are shown. Each redshift range is specified in the legends.}
    \label{fig:chi2_mudif}
\end{figure*}

We also compute the relative standard deviation (Rel.~SD), calculated by taking the median of the absolute values of $\Delta_{\mu}$ and multiplying by 1.48 (an assumed factor for normally distributed data) for all variant redshift sets thus comparing the median absolute deviations to the standard deviation \citep{Hoaglin00}.
To be self-consistent, for both our group analysis and CF analysis, we ensure that, respectively, the same list of SNe (N =~$\NSNeincoherents$ for CFs and N =~$\NSNeingroups$ for groups) are used among variants in the calculation.
The $X^2$ and Rel.~SD values for each of our redshift variants are shown in Fig.~\ref{fig:chi2_mudif} and reported in Tables~\ref{tab:group_results} and~\ref{tab:coherent_results}.

\subsubsection{Group Scatter}\label{sec:group_scatter}
For our multiple methods of determining a group redshift, we compare the $X^2$ of the Hubble residuals of the $\NSNeingroups$ SNe found in any of the group catalogs (Fig.~\ref{fig:chi2_mudif} top left panel; Table~\ref{tab:group_results}) to the $X^2$ of the same sample when the galaxy redshifts are used. 
While using 2M++ C15 corrections, when we replace the galaxy redshifts with the group-averaged redshifts we find $\Delta X^2$ improvements of $92.5$, $63.5$, and $71.9$ for T15, Lim17, and Lam20 respectively. 
We find little to no change in $X^2\ (\Delta X^2 < 1)$ whether the group-averaged or mass-weighted method is employed.  
For T15 and Lim17 group corrections,
averaging within the group reduces scatter compared to taking the redshift of the brightest galaxy.
We find C07 to be significantly less effective at reducing scatter than corrections from all other group catalogs.
Interestingly, we find that applying additional CF corrections from 2M++ C15 improves the $X^2$ for the T15, Lam20, and C07 cases but, for Lim17, the $X^2$ values get worse for all group-averaging methods $(\Delta X^2 < 15)$ compared to not using CF corrections.

As an additional measure of efficacy of the sets of CF corrections and their resulting $X^2$ values, we randomize the corrections applied to each redshift for each variant and calculate the $X^2$ value for the new redshift variant.  We perform this procedure 10 times to determine a distribution of the $X^2$ using randomized values and use this distribution to give a significance of the $\Delta X^2$ of the actual PV correction.
All group-corrected redshift variants have better $X^2$ values than the randomized group corrections. We find that the $X^2$ values for the actual group corrections are all better than the mean $X^2$ values from the randomized corrections by 4.0$\sigma$, 2.4$\sigma$, and 3.1$\sigma$ for T15, Lim17, and Lam20 respectively.

In the second panel on the left in Fig.~\ref{fig:chi2_mudif}, we compare the Rel.~SD values obtained from our group analysis. When 2M++ C15 corrections are applied, the Rel.~SD values follow the same general trend as the $X^2$ values in the panel above, and differences in the trends are likely due to less sensitivity to SNe with larger magnitudes of Hubble residuals than the $X^2$ calculation. Although the C07 corrections without a CF correction give a low Rel.~SD value, we have found C07 to be significantly worse at reducing scatter elsewhere.

\begin{figure*}[!htb]
    \centering
    \includegraphics[width=\textwidth]{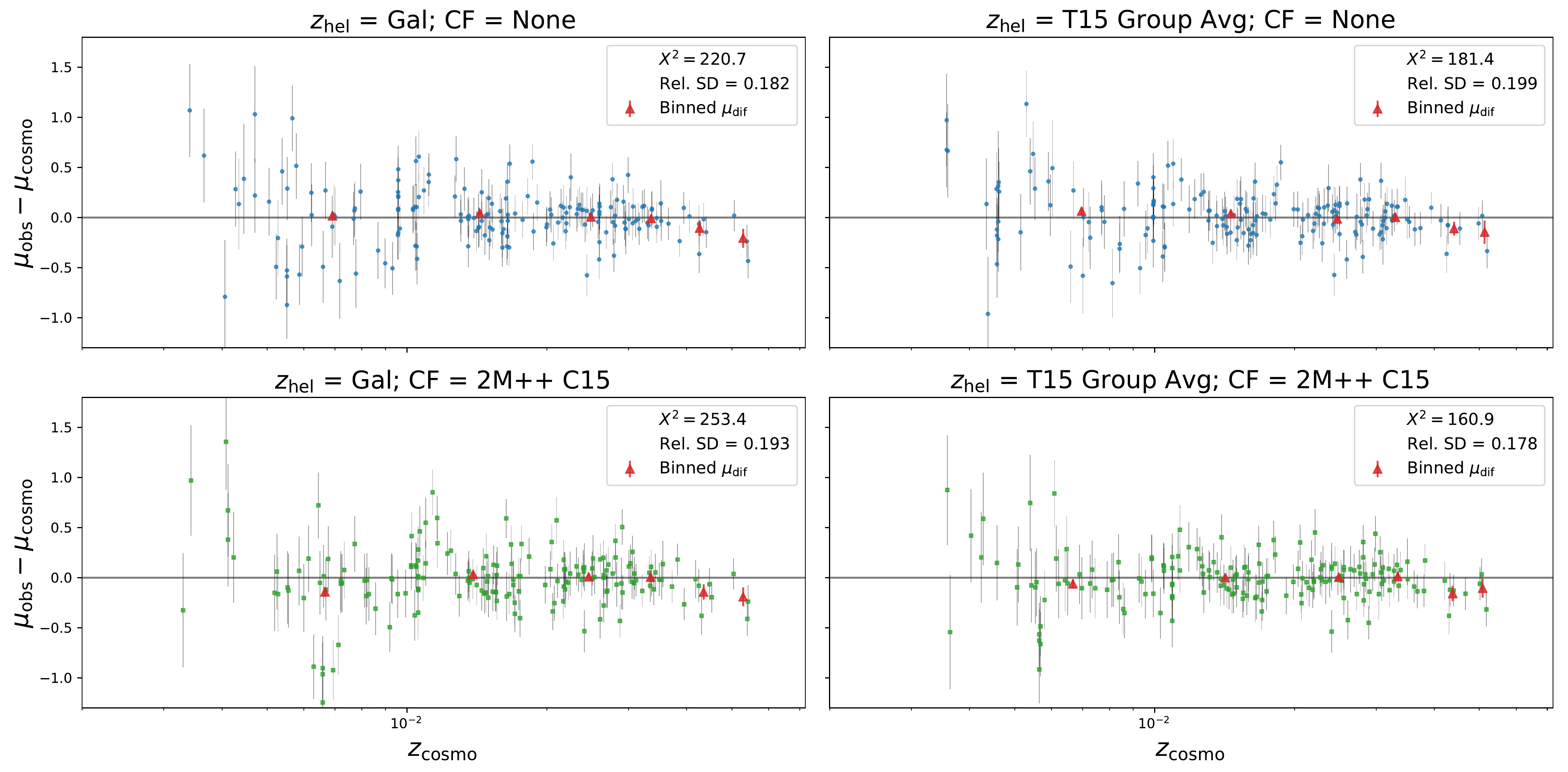}
    \caption{Hubble residuals for $z<0.1$ using the galaxy redshifts and T15 group-corrected redshifts of the SNe found in groups (N = $\NSNeingroups$). Corresponding $X^2$ and Rel.~SD values are given in the legends. Binned (with $\Delta z = 0.01$) $\mu_\textrm{dif}$ values are overplotted. Group corrections visibly reduce the diagonal scatter of these SNe.}
    \label{fig:group_HR_2x2}
\end{figure*}

Taking the group-averaged redshifts from T15 (the best set of group-averaged redshifts among all group catalogs) and taking the galaxy redshifts, we present the Hubble residuals to a best-fit cosmology in Fig.~\ref{fig:group_HR_2x2}. The top two panels contain no CF information, while the bottom two panels use CF corrections from 2M++ C15.  
We find that by including CF corrections, the scatter improves when group information is included ($X^2 = \gTgrnonechi$ with Rel.~SD = $\gTgrnoneRSD$~mag as compared to $X^2 = \gTgrchi$ and Rel.~SD = $\gTgrRSD$~mag). We observe that group averaging improves scatter, but 
by including CF corrections from 2M++ C15 and incorporating group averaging, this improves the $X^2$ even more ($\Delta X^2$ =~\grTdelchi\ improvement from not including either and $\Delta X^2$ = $-92.5$ improvement compared to only using CF corrections).

\begin{deluxetable*}{ccccccc}
\tabletypesize{\footnotesize}
\tablenum{2}
\tablecaption{Group Results (N = \NSNeingroups)}\label{tab:group_results}
\tablewidth{0pt}
\tablehead{\colhead{$z_\textrm{hel}$} & \colhead{CF} & \colhead{$X^2$} & \colhead{Rel.~SD} & \colhead{$\mu_\textrm{dif}$} & \colhead{$\Delta w$} & \colhead{$\Delta H_0$} \\[-0.2cm]
\colhead{} & \colhead{} & \colhead{} & \colhead{(mag)} & \colhead{(mag)} & \colhead{} & \colhead{(km~s$^{-1}$ Mpc$^{-1}$)}}
\decimals
\startdata
Gal & None & \gnonechi & \gnoneRSD & \gnonemu & \gnonew & \gnoneH \\
T15 Group Avg & None & \gTgrnonechi & \gTgrnoneRSD & \gTgrnonemu & \gTgrnonew & \gTgrnoneH \\
T15 Brightest & None & \gTbrnonechi & \gTbrnoneRSD & \gTbrnonemu & \gTbrnonew & \gTbrnoneH \\
T15 Group Mass Weight & None & \gTgmwnonechi & \gTgmwnoneRSD & \gTgmwnonemu & \gTgmwnonew & \gTgmwnoneH \\
Lim17 Group Avg & None & \gLimgrnonechi & \gLimgrnoneRSD & \gLimgrnonemu & \gLimgrnonew & \gLimgrnoneH \\
Lim17 Brightest & None & \gLimbrnonechi & \gLimbrnoneRSD & \gLimbrnonemu & \gLimbrnonew & \gLimbrnoneH \\
Lim17 Group Mass Weight & None & \gLimgmwnonechi & \gLimgmwnoneRSD & \gLimgmwnonemu & \gLimgmwnonew & \gLimgmwnoneH \\
Lam20 Group Avg & None & \gLamgrnonechi & \gLamgrnoneRSD & \gLamgrnonemu & \gLamgrnonew & \gLamgrnoneH \\
C07 Group Avg & None & \gCgrnonechi & \gCgrnoneRSD & \gCgrnonemu & \gCgrnonew & \gCgrnoneH \\
\hline
Gal & 2M++ C15 & \gchi & \gRSD & \gmu & \gw & \gH \\
T15 Group Avg & 2M++ C15 & \gTgrchi & \gTgrRSD & \gTgrmu & \gTgrw & \gTgrH \\
T15 Brightest & 2M++ C15 & \gTbrchi & \gTbrRSD & \gTbrmu & \gTbrw & \gTbrH \\
T15 Group Mass Weight & 2M++ C15 & \gTgmwchi & \gTgmwRSD & \gTgmwmu & \gTgmww & \gTgmwH \\
Lim17 Group Avg & 2M++ C15 & \gLimgrchi & \gLimgrRSD & \gLimgrmu & \gLimgrw & \gLimgrH \\
Lim17 Brightest & 2M++ C15 & \gLimbrchi & \gLimbrRSD & \gLimbrmu & \gLimbrw & \gLimbrH \\
Lim17 Group Mass Weight & 2M++ C15 & \gLimgmwchi & \gLimgmwRSD & \gLimgmwmu & \gLimgmww & \gLimgmwH \\
Lam20 Group Avg & 2M++ C15 & \gLamgrchi & \gLamgrRSD & \gLamgrmu & \gLamgrw & \gLamgrH \\
C07 Group Avg & 2M++ C15 & \gCgrchi & \gCgrRSD & \gCgrmu & \gCgrw & \gCgrH \\
\enddata
\tablecomments{For each of the group variants detailed in Table~\ref{tab:group} and using multiple methods of determining the group redshift in section~\ref{subsec:group_utilize}, we present the $X^2$ and Rel.~SD values of the group corrections. Furthermore we show the impact of these corrections on the change in mean distance modulus $\mu_\textrm{dif}$ as well as on the measured cosmological parameters $w$ and $H_0$. We present an additional set of results when we apply CF corrections using 2M++ C15. Here we assume $\sigma_\textrm{int} = \groupsigint$ mag.}
\end{deluxetable*}

\begin{deluxetable*}{ccccccc}
\tabletypesize{\footnotesize}
\tablenum{3}
\tablecaption{Coherent-Flow Results (N = \NSNeincoherents)}\label{tab:coherent_results}
\tablewidth{0pt}
\tablehead{\colhead{$z_\textrm{hel}$} & \colhead{CF} & \colhead{$X^2$} & \colhead{Rel.~SD} & \colhead{$\mu_\textrm{dif}$} & \colhead{$\Delta w$} & \colhead{$\Delta H_0$} \\[-0.2cm]
\colhead{} & \colhead{} & \colhead{} & \colhead{(mag)} & \colhead{(mag)} & \colhead{} & \colhead{(km~s$^{-1}$ Mpc$^{-1}$)}}
\decimals
\startdata
Gal & None & \bnonechi & \bnoneRSD & \bnonemu & \bnonew & \bnoneH \\
Gal & 2M++ C15 & \bfullchi & \bfullRSD & \bfullmu & \bfullw & \bfullH \\
Gal & $-$2M++ C15 & \bnegfullchi & \bnegfullRSD & \bnegfullmu & \bnegfullw & \bnegfullH \\
Gal & 2M++$_\textrm{ilos}$ & \bdarcychi & \bdarcyRSD & \bdarcymu & \bdarcyw & \bdarcyH \\
Gal & 2M++ [Gr] & \bgrchi & \bgrRSD & \bgrmu & \bgrw & \bgrH \\
Gal & 2M++/SDSS & \bsdsschi & \bsdssRSD & \bsdssmu & \bsdssw & \bsdssH \\
Gal & 2M++/SDSS$_\textrm{ilos}$ & \bsdssdarcychi & \bsdssdarcyRSD & \bsdssdarcymu & \bsdssdarcyw & \bsdssdarcyH \\
Gal & 2M++/SDSS/6dF & \bsixdfchi & \bsixdfRSD & \bsixdfmu & \bsixdfw & \bsixdfH \\
Gal & 2M++$_\textrm{ne}$ & \bnechi & \bneRSD & \bnemu & \bnew & \bneH \\
Gal & \textit{Cf3} & \bcfchi & \bcfRSD & \bcfmu & \bcfw & \bcfH \\
Gal & $-$\textit{Cf3} & \bnegcfchi & \bnegcfRSD & \bnegcfmu & \bnegcfw & \bnegcfH \\
Gal & VGAS & \bvgaschi & \bvgasRSD & \bvgasmu & \bvgasw & \bvgasH \\
Gal & $-$VGAS & \bnegvgaschi & \bnegvgasRSD & \bnegvgasmu & \bnegvgasw & \bnegvgasH \\
Gal & 2MRS & \btmrschi & \btmrsRSD & \btmrsmu & \btmrsw & \btmrsH \\
Gal & 2MRS$_\textrm{ilos}$ & \btmrsiloschi & \btmrsilosRSD & \btmrsilosmu & \btmrsilosw & \btmrsilosH \\
\hline
Gal + T15 Group Avg & None & \bTnonechi & \bTnoneRSD & \bTnonemu & \bTnonew & \bTnoneH \\
Gal + T15 Group Avg & 2M++ C15 & \bTfullchi & \bTfullRSD & \bTfullmu & \bTfullw & \bTfullH \\
Gal + T15 Group Avg & 2M++$_\textrm{ilos}$ & \bTdarcychi & \bTdarcyRSD & \bTdarcymu & \bTdarcyw & \bTdarcyH \\
Gal + T15 Group Avg & 2M++ [Gr] & \bTgrchi & \bTgrRSD & \bTgrmu & \bTgrw & \bTgrH \\
Gal + T15 Group Avg & 2M++/SDSS & \bTsdsschi & \bTsdssRSD & \bTsdssmu & \bTsdssw & \bTsdssH \\
Gal + T15 Group Avg & 2M++/SDSS$_\textrm{ilos}$ & \bTsdssdarcychi & \bTsdssdarcyRSD & \bTsdssdarcymu & \bTsdssdarcyw & \bTsdssdarcyH \\
Gal + T15 Group Avg & 2M++/SDSS/6dF & \bTsixdfchi & \bTsixdfRSD & \bTsixdfmu & \bTsixdfw & \bTsixdfH \\
Gal + T15 Group Avg & 2M++$_\textrm{ne}$ & \bTnechi & \bTneRSD & \bTnemu & \bTnew & \bTneH \\
Gal + T15 Group Avg & \textit{Cf3} & \bTcfchi & \bTcfRSD & \bTcfmu & \bTcfw & \bTcfH \\
Gal + T15 Group Avg & VGAS & \bTvgaschi & \bTvgasRSD & \bTvgasmu & \bTvgasw & \bTvgasH \\
Gal + T15 Group Avg & 2MRS & \bTtmrschi & \bTtmrsRSD & \bTtmrsmu & \bTtmrsw & \bTtmrsH \\
Gal + T15 Group Avg & 2MRS$_\textrm{ilos}$ & \bTtmrsiloschi & \bTtmrsilosRSD & \bTtmrsilosmu & \bTtmrsilosw & \bTtmrsilosH \\
\enddata
\tablecomments{Similar to Table \ref{tab:group_results} we present the full results from our CF analysis.
For the variants labeled with $z_\textrm{hel}$ = Gal + T15 Group Avg, redshifts are updated according to the T15 group-averaged redshifts for only the $\NSNeingroups$ SNe in groups and redshifts are left with $z_\textrm{hel}$ =~Gal otherwise. The different CF variants are explained in section~\ref{subsec:CF_variants}. We use $\sigma_\textrm{int} = \coherentsigint$ mag.}
\end{deluxetable*}

\subsubsection{Coherent-Flow Scatter}\label{sec:coherent-flow_scatter}
Similarly, we measure the improvement in Hubble residuals from using the many different models for CFs: 2M++, \textit{Cf3}, VGAS, and 2MRS from section~\ref{subsec:CF_variants} along with their variants. Results for all variants are given in Table~\ref{tab:coherent_results}. In the top right panel of Fig.~\ref{fig:chi2_mudif}, we show the $X^2$ of the sample of $\NSNeincoherents$ SNe when a select number of these models are used.

Since the sign of CF corrections has often been confused in the community, we flip the sign of each set of corrections by $-1$ and test them to ensure we avoid this pitfall. 
We see that for all the cases, reversing the sign of the corrections makes the $X^2$ worse.

As an additional test, for the 2M++ C15 model, we also scaled the predicted velocity corrections by $50\%$ ($0.50 \times \textrm{2M++ C15}$) and found that doing so results in an improvement in the scatter.
This finding is reinforced by the reduction in scatter observed from the 2M++/SDSS and 2M++/SDSS/6dF variants, which have smaller values for $\beta$ (0.314 and 0.341 respectively) than that from 2M++ C15 (0.431).

Using the group center to calculate the predicted CF correction (the 2M++ [Gr] variant) results in a large improvement in the scatter ($\Delta X^2 = -46.5$ compared to 2M++ C15 when group-corrected redshifts are not also included), but integrating over all possible distances along the line of sight (2M++$_\textrm{ilos}$) results in an even larger improvement in scatter ($\Delta X^2~=~-66.0$). The 2M++/SDSS, 2M++/SDSS/6dF, and 2MRS variants are similar in scatter improvement with $\Delta X^2$ ranging from $\sim-35$--$45$.

Using the randomization method as described above, we find that CF~=~2M++~C15 improves the $X^2$ by 4.3$\sigma$ from the mean of the set of randomized CFs.
This suggests that, while the 2M++ C15 corrections (without group-corrected redshifts) do not improve the total $X^2$, they are better than applying random corrections of a similar magnitude.
Furthermore, considering that we use roughly three times as many SNe in the CF analysis as in the group analysis, it is noteworthy that none of the CFs improve the $X^2$ as much as the group corrections.  
Similar to the group Rel.~SD panel on the left, the CF Rel.~SD values as seen in the right side of Fig.~\ref{fig:chi2_mudif} follow the same general trend as the $X^2$ values.

Additionally, we measure the impact of the CF corrections when we also include group averaging from T15 ($z_\textrm{hel}$ = Gal + T15 Group Avg). As shown in the top right panel of Fig.~\ref{fig:chi2_mudif}, there is a large improvement in $X^2$ for all variants when groups are incorporated as well.
When including group-corrected redshifts, the 2M++/SDSS$_\textrm{ilos}$ variant results in the largest improvement in scatter among CF corrections as compared to 2M++ C15 ($\Delta X^2~=~-39.0$), but the overall improvement in scatter is comparable to the improvements seen from the 2M++$_\textrm{ilos}$, 2M++/SDSS, 2M++/SDSS/6dF, 2MRS, and 2MRS$_\textrm{ilos}$ variants.
Similar to section~\ref{sec:group_scatter}, to better understand the relative contributions from group corrections and CF corrections, we present the contributions to the total $X^2$ values from various sets of SNe within our 2M++ CF analysis in Table~\ref{tab:coherent_in_groups} and Table~\ref{tab:coherent_not_in_groups}.  We see that while group assignment reduces the $X^2$, the largest reduction is when both group corrections and CF corrections are applied.

\begin{deluxetable}{cccccc}
\tabletypesize{\footnotesize}
\tablenum{4}
\tablecaption{SNe in Groups (N = \NSNeingroups)}\label{tab:coherent_in_groups}
\tablewidth{0pt}
\tablehead{\colhead{$z_\textrm{hel}$} & \colhead{CF} & \colhead{$X^2$} & \colhead{$\Delta X^2$} & \colhead{Reduced $X^2$} & \colhead{Rel.~SD} \\[-0.2cm]
\colhead{} & & & & & \colhead{(mag)}}
\decimals
\startdata
Gal & None & \grindivnonechi & \grindivnonedelchi & \grindivnoneredchi & \grindivnonersd\\
Gal & 2M++ C15 & \grindivchi & \grindivdelchi & \grindivredchi & \grindivrsd\\
T15 Group Avg & None & \grTnonechi & \grTnonedelchi & \grTnoneredchi & \grTnonersd\\
T15 Group Avg & 2M++ C15 & \grTchi & \grTdelchi & \grTredchi & \grTrsd\\
\enddata
\tablecomments{These are values from our CF analysis for those SNe also found in our group analysis. The mean redshift of this sample is $z = 0.018$. 
Summing the $X^2$ values from this table with those from Table~\ref{tab:coherent_not_in_groups} results in values presented in Table~\ref{tab:coherent_results}.}
\end{deluxetable}

\begin{deluxetable}{cccccc}
\tabletypesize{\footnotesize}
\tablenum{5}
\tablecaption{SNe not in Groups (N = \NSNenotingroups)}\label{tab:coherent_not_in_groups}
\tablewidth{0pt}
\tablehead{\colhead{$z_\textrm{hel}$} & \colhead{CF} & \colhead{$X^2$} & \colhead{$\Delta X^2$} & \colhead{Reduced $X^2$} & \colhead{Rel.~SD}\\[-0.2cm]
\colhead{} & & & & & \colhead{(mag)}}
\decimals
\startdata
Gal & None & \ngrindivnonechi & \ngrindivnonedelchi & \ngrindivnoneredchi & \ngrindivnonersd\\
Gal & 2M++ C15 & \ngrindivchi & \ngrindivdelchi & \ngrindivredchi & \ngrindivrsd\\
\enddata
\tablecomments{These are values from our CF analysis for those SNe that are specifically not found in any of the group catalogs. The mean redshift of this sample is $z = 0.033$.}
\end{deluxetable}

\begin{figure*}[!htb]
    \centering
    \includegraphics[width=\textwidth]{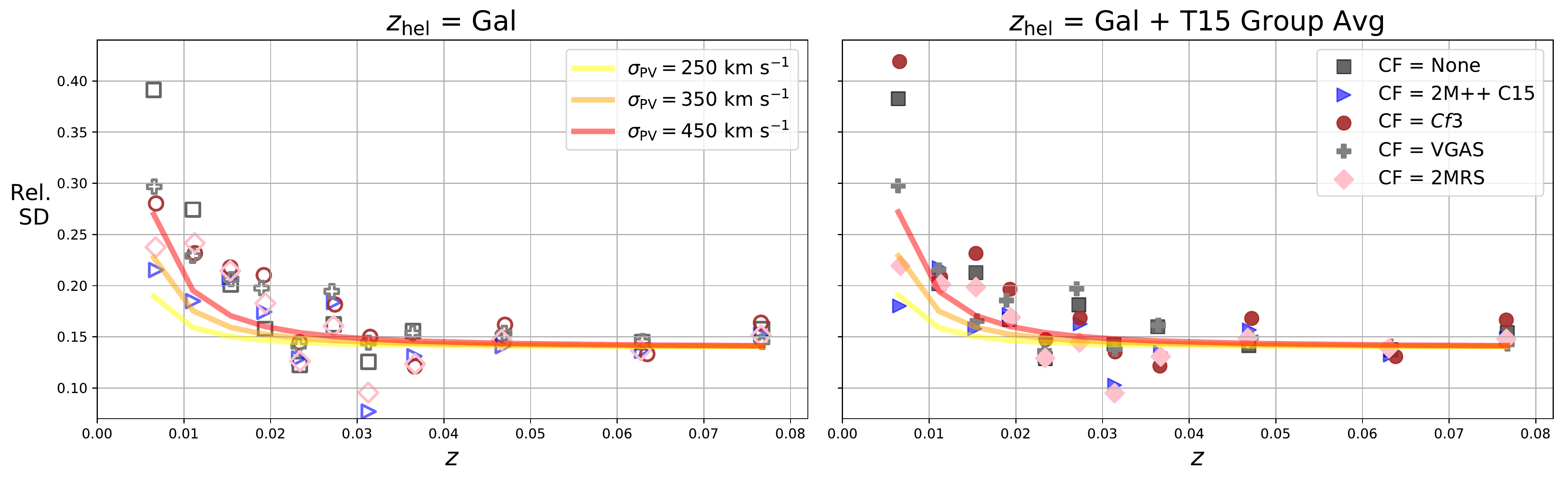}
    \caption{Binned with equal SNe per bin Rel.~SD values from our CF analysis as compared to CF~=~None. Overplotted on the panels are the expected Rel.~SD values for three different values of PV uncertainties. \textbf{Left}: All SNe used for CF analysis (N =~$\NSNeincoherents$). \textbf{Right}: Heliocentric redshifts for SNe found in groups are updated according to T15 (N =~$\NSNeincoherents$ but $\NSNeingroups$ SN redshifts are group corrected as well).}
    \label{fig:coherent_zbin}
\end{figure*}

We also study the relative contributions of the improvements to the Hubble diagram scatter in redshift bins with equal numbers of SNe per bin and present comparative Rel.~SD values to CF~=~None in Fig.~\ref{fig:coherent_zbin}. The left panel of Fig.~\ref{fig:coherent_zbin} is calculated using all SNe used in our CF analysis, while the right panel has those SNe in groups updated according to the group-averaged T15 redshifts where available. 
We find CF corrections to have a more significant impact at lower redshifts, as expected, but we find no trends of reduction in scatter comparing the left panel to the right panel otherwise.
Following \cite{Scolnic18}, we also try to use the dispersion in redshift bins to determine the PV uncertainty.  We overplot the impact of uncertainties of 250, 350, and 450 km~s$^{-1}$. 
While we observe the expected trend of Rel.~SD versus redshift, we cannot constrain the PV uncertainty with this type of analysis beyond $\sim100$ km~s$^{-1}$.

\subsection{Impact on Measurement of Cosmological Parameters}\label{subsec:impact_cosmo_params}
We define $\mu_\textrm{dif}$ for each sample as the difference in the weighted mean of the low-redshift SN vector of $\Delta_\mu$ (Eq.~\ref{eq:mu_dif}) with uncertainties $\sigma_\mu$ (Eq.~\ref{eq:distance_error}) and plot $\mu_\textrm{dif}$ in the third panels of Fig.~\ref{fig:chi2_mudif}.
We propagate these differences in $\mu$ to a difference in both $w$ and $H_0$ in the fourth and fifth panels of Fig.~\ref{fig:chi2_mudif}.
Following \citet{KesslerScolnic17}, we measure the change in $w$ using the \textbf{wfit} program in \texttt{SNANA}, with approximate priors from Planck CMB measurements \citep{Planck18} for $z < 2.2$ (high-$z$ SNe presented in Pantheon+).
To measure the change in $H_0$, we determine the change in the mean intercept, as shown in the bottom panel of Fig.~\ref{fig:chi2_mudif}, but limited to a redshift range of $0.023<z<0.15$ and follow the formulas outlined in \citet{Riess16}. These values are also given in Tables~\ref{tab:group_results}~and~\ref{tab:coherent_results}.

\subsubsection{Impact from Group Corrections}
In the third panel on the left of Fig.~\ref{fig:chi2_mudif}, we show the change in the mean value of the distance modulus residuals $\mu_\textrm{dif}$ for all the SNe impacted by the group corrections to redshifts. 
We find that the impact of the group corrections is on the few millimagnitudes scale when only looking at the $\NSNeingroups$ SNe affected, which would be on the 1 mmag scale when including the full set of SNe at low~$z$.

For $w$, we compare the same galaxy redshift variant to the unweighted-averaged variants using the whole redshift range from $z<2.2$ and find $\sigma_{w}= 0.005$, which is again a small amount relative to the statistical uncertainty of $0.04$ \citep[fourth panel on the left of Fig.~\ref{fig:chi2_mudif}; ][]{Scolnic18}.
As for $H_0$, all changes are within $\sigma_{\textrm{H}_0}=0.05$ km~s$^{-1}$ Mpc$^{-1}$, which is a negligible fraction of the uncertainty in $H_0$ of $\sim$ 1.5 km~s$^{-1}$ Mpc$^{-1}$ \citep[fifth panel on the left of Fig.~\ref{fig:chi2_mudif}; ][]{Riess19}.

\subsubsection{Impact from Coherent-Flow Corrections}
We make the same assessments for the CF corrections.  The third panel on the right of Fig.~\ref{fig:chi2_mudif} presents the mean distance modulus residuals in comparison to the CF~=~None redshift variant $\mu_\textrm{dif}$ for SNe with $z < 0.08$. Overall, the impact on $\mu_\textrm{dif}$ is much larger for the CF corrections than the group corrections. 
Comparing CF~=~2M++ C15 to CF~=~None we observe a $\Delta\mu_\textrm{dif}$ of \bfullmu\ mag. The relative difference between CF~=~\textit{Cf3} and CF~=~2M++ C15 is $\Delta\mu_\textrm{dif} = -0.038$ mag, which we find to be largely due to the velocity offset of 91 km~s$^{-1}$ shown in Fig.~\ref{fig:cf3_vpec_residual}. 
Comparing CF~=~2MRS to CF = 2M++ C15, $\Delta\mu_\textrm{dif} = 0.011$ mag.

From the fourth and fifth panels on the right of Fig.~\ref{fig:chi2_mudif} and as seen in Table~\ref{tab:coherent_results}, we measure a change of $\Delta w=$~\bfullw\ and $\Delta \textrm{H}_0=$~\bfullH~km~s$^{-1}$ Mpc$^{-1}$ between the CF~=~2M++ C15 and CF~=~None cases.  
The largest disparities are seen when using the \textit{Cf3} and VGAS corrections, however as seen in the $X^2$ and Rel.~SD panels, both \textit{Cf3} and VGAS also make the scatter worse by the largest amounts.
Between the various CF treatments of CF~=~2M++~C15, CF~=~2M++$_\textrm{ilos}$, CF~=~2M++~[Gr], CF~=~2M++/SDSS, and CF~=~2M++/SDSS/6dF, $w$ differs by no more than 0.02 and $H_0$ differs by no more than 0.13~km~s$^{-1}$ Mpc$^{-1}$ because the mean velocities of the sets of corrections are all within 12 km~s$^{-1}$.
The CF treatments that result in the greatest improvement in scatter, CF~=~2M++/SDSS$_\textrm{ilos}$ and CF~=~2MRS, vary $w$ by 0.04 and $H_0$ by 0.18~km~s$^{-1}$ Mpc$^{-1}$.
Comparing CF~=~2M++/SDSS$_\textrm{ilos}$ and CF~=~2MRS$_\textrm{ilos}$, $w$ varies by 0.01 and $H_0$ varies by 0.08~km~s$^{-1}$ Mpc$^{-1}$.
The actual uncertainty in $H_0$ and $w$ for future analyses is discussed in the next section.

\section{Discussion}\label{sec:Discussion}
\subsection{What to Do For Upcoming Cosmological Analyses}

In this paper, we study two corrections due to PVs for low-redshift galaxies. The first correction is due to assigning galaxies to groups. We examine multiple different methods and find a relatively similar impact.
The choice of using group redshifts is clear, and the data prefers the T15 catalog, with the second best option being Lam20 which can be used as a systematic variant.

For CF corrections, we find that the 2M++ and 2MRS corrections are optimal. However, from Fig.~\ref{fig:chi2_mudif}, we show that in terms of reducing the $X^2$, integrating over all possible distances along the line of sight and updating $\beta$ and the $v_\textrm{ext. coh.}$ according to SDSS and 2M++ data, or using the 2MRS corrections is preferred. 
Past analyses by SN cosmologists measuring $w$ \citep{Betoule14,Scolnic18,Brout19} have not used these techniques.
One potential explanation for the improvement observed from integrating along the line of sight is that this integration helps account for noise at small scales without removing large-scale correlations that affect the central values of $w$/$H_0$. Group corrections however remove nonlinear motions that cause large scatter at low redshift, but have little effect on those same central values (see Fig.~\ref{fig:graphic} for visualization).  Furthermore, in Fig.~\ref{fig:chi2_mudif}, we show how a larger redshift range of Pantheon+ can help further discriminate between different PV models. 
One possible future analysis would be to include a PV offset for $z<0.06$ as an additional parameter in the cosmological fit; we have seen doing so has little effect on $H_0$ ($<0.2$), but a larger effect on $w$ ($>0.05$).

These findings have implications for other measurements on the local distance ladder as well, besides SH0ES. Both \citet{Burns18} and \citet{Freedman19} provide values for $H_0$ and use the CSP SN redshifts in the CMB frame. When we limit our sample to only those SNe from CSP, we find including the CF corrections raises $H_0$ by 0.4--0.5 km~s$^{-1}$ Mpc$^{-1}$ while improving the $X^2$ (reduced $X^2$ goes from 1.02 to between 0.92 and 0.98).
Our findings agree with the results of \citet{Sedgwick21} who find that accounting for environmentally induced PVs of SN Ia host galaxies does not resolve the Hubble tension. Additionally, we note that PVs cannot affect measurements at high $z$, which can be used to trace the Hubble parameter \citep{Dainotti21}.

\subsection{Other Uses of Redshift Corrections in Cosmological Probes}

While we focus on PVs as a systematic uncertainty in measurements of $H_0$ and $w$ here, the methodology can be inverted and SN PVs can be used as a probe of the growth-of-structure \citep{Gordon07,Johnson14,Carrick15,Castro16,Howlett17gos,Stahl21}. This has been demonstrated with real data by \cite{Boruah20a} and is simulated for the Legacy Survey of Space and Time (LSST) in \cite{Howlett17gos}, \cite{Kim19}, and \cite{Kim20}. Here, the scatter of Hubble residuals is being used to determine the PVs, rather than the PVs being used to reduce the Hubble residual scatter. This probe can be as constraining of growth rate of structure as weak lensing studies at low $z$ \citep{Boruah20a}.

We note that, while the effects on $H_0$ and $w$ are on the 1\% level, these issues have a much larger effect when measuring $H_0$ with megamasers \citep{Pesce20}, kilonovae  \citep{Howlett19,Nicolaou20,Mukherjee21}, or gravitational wave events \citep{Fishbach19}. For these, the number of objects are $<10$ and $z\sim0.02$, so uncertainty in the PV corrections can be as large as the uncertainty in the distance measurement \citep{Howlett19}. Cross-checking PV treatments with larger low-$z$ SNe samples will be very valuable for these studies.  

\section{Conclusions}\label{sec:Conclusions}
We study two redshift correction methods due to PVs for low-redshift galaxies. The first is incorporating galaxy groups. The second is applying CF corrections. We find that the strongest improvement in the scatter of Hubble diagram residuals are when host galaxies of SNe can be assigned to groups of galaxies and CF corrections are applied on the groups.  We determine the optimal method of group assignment is presented in T15 and the optimal method for CF corrections is given using 2M++/SDSS$_\textrm{ilos}$ or 2MRS$_\textrm{ilos}$. The optimal PV corrections can be constrained such that the impact on $H_0$ is $0.06$--$0.11$ km~s$^{-1}$ Mpc$^{-1}$ and the impact on $w$ is $0.02$--$0.03$.

The data set analyzed in this paper, which is released in \citet{Carr21}, will remain valuable for years, as for  measurements  of  $H_0$,  the  current  low-$z$ data set is likely irreplaceable because it contains measurements of SNe used in the second rung of the distance ladder, and roughly one of these is discovered per year. Therefore, further study on this set and these corrections is encouraged.  
Particularly, spectroscopic programs should 
continue to grow the number of redshift identifications in low-redshift galaxies
to improve the low yield for galaxies in groups.

\begin{acknowledgements}
We thank the reviewers for their expeditious and thorough reviews of our paper. We thank Robert Lilow and Adi Nusser for their expertise with 2MRS and the valuable information they provided to us.
We also thank Cullan Howlett for insightful discussions.
This work was also supported by resources provided by the University of Chicago Research Computing Center.
D.S.~is supported by DOE grant DE-SC0010007 and the David and Lucile Packard Foundation. D.S.~and B.M.R.~are supported in part by the National Aeronautics and Space Administration (NASA) under Contract No.~NNG17PX03C issued through the Roman Science Investigation Teams Programme. 
A.G.R.~gratefully acknowledges support by the Munich Institute for Astro- and Particle Physics (MIAPP) of the DFG cluster of excellence “Origin and Structure of the Universe.”
D.B.~acknowledges support for this work was provided by NASA through the NASA Hubble Fellowship grant HSTHF2-51430.001 awarded by the Space Telescope Science Institute, which is operated by the Association of Universities for Research in Astronomy, Inc., for NASA, under contract NAS5-26555. 
T.D., A.C., and K.S.~gratefully acknowledge Australian Research Council’s Laureate Fellowship (project FL180100168). 
D.O.J.~is supported by a Gordon and Betty Moore Foundation postdoctoral fellowship at the University of California, Santa Cruz.  Support for this work was provided by NASA through the NASA Hubble Fellowship grant HF2-51462.001 awarded by the Space Telescope Science Institute, which is operated by the Association of Universities for Research in Astronomy, Inc., for NASA, under contract NAS5-26555.

Software: 
\texttt{SNANA} \citep{Kessler09}, {astropy} \citep{astropy:2013,astropy:2018}
{matplotlib} \citep{Hunter07},
{numpy} \citep{numpy11}, {PIPPIN} \citep{Pippin}.
\end{acknowledgements}

\bibliographystyle{mn2e}
\bibliography{main}{}

\appendix
\section{Propagation of PV uncertainty to Cosmological Redshift }\label{sec:variablevariance}
\begin{figure}
    \centering
    \includegraphics[width=12cm]{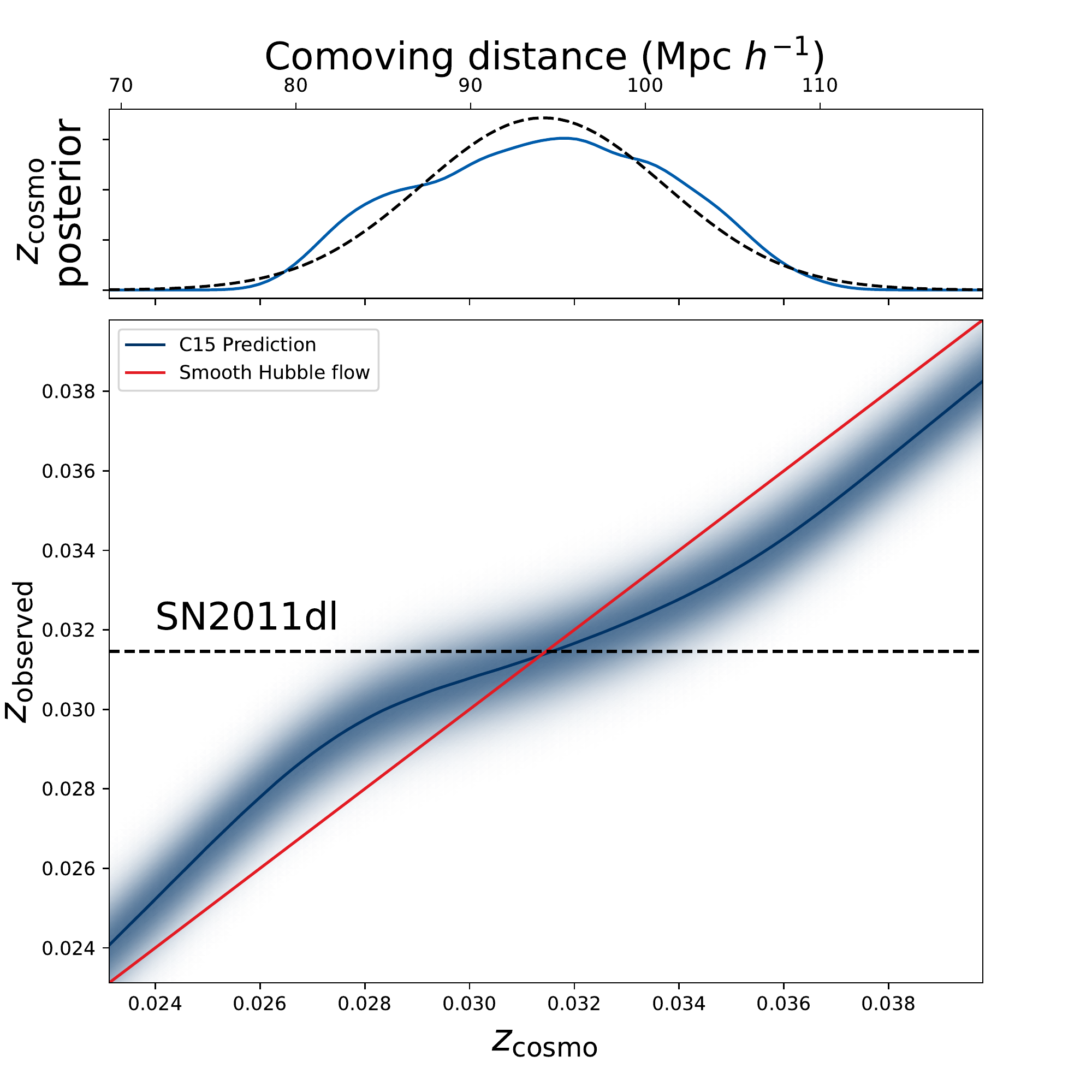}
    \caption{In the lower panel, we show the relation between cosmological and observed redshift predicted by the \citet{Carrick15} reconstruction along the line of sight for the supernova 2011dl. The shading about the predicted relation shows the expected scatter with 250 km~s$^{-1}$ of PV dispersion. The distribution of the underlying cosmological redshift can be found by integrating along the line at the observed redshift ($z_\textrm{CMB}=0.031459$); the upper panel shows the sampled posterior probability density function (PDF). As this SN is located near a predicted galaxy overdensity, the relation between predicted observed redshift and comoving distance/cosmological redshift takes on a characteristic $S$ shape. Larger density concentrations can cause this curve to become non-monotonic, and the region becomes triply valued. A \SI{250}{\kilo\meter\per\second} uncertainty in PV dispersion here becomes a broader \SI{700}{\kilo\meter\per\second} uncertainty in cosmological redshift; there are many underlying cosmological redshifts potentially consistent  with the observation. The PDF can be seen to be non-Gaussian where the second derivative of the relation is high, however we approximate it as Gaussian for computational reasons. }
    \label{fig:reconstructemapping}
\end{figure}

To this point, in order to compare PV catalogs using the $X^2$ statistic, we have fixed the assumed uncertainties in the PV corrected redshifts to \SI{250}{\kilo\meter\per\second} for all SNe and for all redshift vectors. However when making use of PV corrections based on an underlying reconstruction such as those of \citet{Carrick15} and \citet{LilowNusser21}, a constant peculiar velocity uncertainty can translate into a variable uncertainty in the cosmological redshift. This effect can easily be seen by reference to `triple-valued regions'; when looking along the line of sight through a sufficiently overdense collapsing volume, there are three distinct positions along the line of sight that produce a single observed redshift. Given an observed redshift along such a line of sight, even without any nonlinear dispersion in PVs, the underlying cosmological redshift is uncertain between three potential values. In practice, PV reconstructions are smoothed, and only the densest concentrations of mass in the nearby universe (e.g.,~the Coma cluster) will be modeled as genuinely triple-valued regions. However the cosmological redshift uncertainty will vary from SN to SN depending on their position in the reconstructed density/velocity fields. We illustrate an example using the supernova 2011dl in Fig.~\ref{fig:reconstructemapping}.

The integrated line-of-sight method provides a posterior distribution over the cosmological redshift, allowing us to estimate both the central value and dispersion of the cosmological redshift. We use the same methodology as the 2M++$_\textrm{ilos}$ variant discussed in section \ref{subsec:CF_variants} to sample from the posterior. As the reconstructions are smoothed, we approximate the distributions of the cosmological redshifts as Gaussian, and take the posterior mean of the redshift as our redshift vector and the posterior variance as a redshift uncertainty. This approach is similar to the one discussed in \citet{Rahman21}.

As a first check on whether these variance estimates are informative about the scatter in the SN\,Ia sample, we compare the Rel.~SD of SNe with high and low estimated cosmological redshift uncertainties. Cutting on median redshift ($z<0.0282$), we observe a Rel.~SD of $0.124$ for SNe with $cz_\textrm{cosmo}$ uncertainties $< 250$ km~s$^{-1}$ but a Rel.~SD of $0.142$ for SNe with $cz_\textrm{cosmo}$ uncertainties $> 250$ km~s$^{-1}$. For $z>0.0282$, we observe a Rel.~SD of $0.121$ for low  uncertainties and a Rel.~SD of $0.129$ for high uncertainties. This indicates that the estimates of the uncertainty due to PV dispersion uncertainties are informative about the Hubble scatter of our sample (greater uncertainties for SNe contributing more scatter).

While a direct comparison of $X^2$ statistics is inappropriate to compare different estimates of the uncertainties, we can instead compare `effective $X^2$' differences based on the definition of a Gaussian log-likelihood, which penalizes the overestimation of uncertainties through the inclusion of a Gaussian normalization term. Based on the definitions given for Eq.~\ref{eq:chi2}, our  $X^2_\textrm{eff}$ is defined
\begin{equation}\label{eq:chi2eff}
    X^2_\textrm{eff} =  \sum_i{2 \log(\sigma_{\mu,i}) + \frac{\Delta_{\mu,i}^2}{\sigma_{\mu,i}^2}},
\end{equation}
where the values of $\sigma_{\mu}$ have been recalculated based on the revised uncertainties in the cosmological redshift.

As can be seen from Table~\ref{tab:chi2_effs}, for most instances comparing fixed PV uncertainties to variable PV uncertainties, the $X^2_\textrm{eff}$ is moderately affected ($X^2_\textrm{eff}$ values differing by $< 10$). Comparing `Fixed' and `Variable' for 2MRS$_\textrm{ilos}$, the $X^2_\textrm{eff}$ gets worse whether or not group redshifts are used. The best $X^2_\textrm{eff}$ is obtained with group redshifts, 2M++/SDSS$_\textrm{ilos}$ (reduced $\beta$ and integration along the line of sight), and fixed PV uncertainties.
Treating PV uncertainties variably for this variant makes the $X^2_\textrm{eff}$ worse by $7.6$. We speculate that as group corrections will reduce the underlying PV scatter in overdense regions, the uncertainties in these redshifts are overestimated, increasing the effective $X^2$. These results show that a more robust treatment of the correspondence between PV uncertainty and Hubble scatter should be incorporated in future analyses.

\newcommand{\bfullchieff}{$-1287.7$}
\newcommand{\bTfullchieff}{$-1381.1$}
\newcommand{\bsdsschieff}{$-1330.5$}
\newcommand{\bTsdsschieff}{$-1408.0$}
\newcommand{\bsdssdarcychieff}{$-1352.9$}
\newcommand{\bTsdssdarcychieff}{$-1420.1$}
\newcommand{\bsdssdarcyvarchieff}{$-1362.5$}
\newcommand{\bTsdssdarcyvarchieff}{$-1412.5$}
\newcommand{\btmrschieff}{$-1334.6$}
\newcommand{\bTtmrschieff}{$-1409.4$}
\newcommand{\btmrsiloschieff}{$-1332.7$}
\newcommand{\bTtmrsiloschieff}{$-1396.9$}
\newcommand{\btmrsilosvarchieff}{$-1323.0$}
\newcommand{\bTtmrsilosvarchieff}{$-1366.4$}

\newcommand{\bfulldelchieff}{$+0.0$}
\newcommand{\bTfulldelchieff}{$-93.4$}
\newcommand{\bsdssdelchieff}{$-42.8$}
\newcommand{\bTsdssdelchieff}{$-120.3$}
\newcommand{\bsdssdarcydelchieff}{$-65.3$}
\newcommand{\bTsdssdarcydelchieff}{$-132.5$}
\newcommand{\bsdssdarcyvardelchieff}{$-74.8$}
\newcommand{\bTsdssdarcyvardelchieff}{$-124.8$}
\newcommand{\btmrsdelchieff}{$-46.9$}
\newcommand{\bTtmrsdelchieff}{$-121.7$}
\newcommand{\btmrsilosdelchieff}{$-45.1$}
\newcommand{\bTtmrsilosdelchieff}{$-109.3$}
\newcommand{\btmrsilosvardelchieff}{$-35.3$}
\newcommand{\bTtmrsilosvardelchieff}{$-78.8$}

\begin{deluxetable}{ccccc}
\tabletypesize{\footnotesize}
\tablenum{A}
\tablecaption{Effective $X^2$ Results}\label{tab:chi2_effs}
\tablewidth{0pt}
\tablehead{\colhead{$z_\textrm{hel}$} & \colhead{CF} & \colhead{PV Uncertainty} & \colhead{$X^2_\textrm{eff}$} & \colhead{$\Delta X^2_\textrm{eff}$} \\}
\decimals
\startdata
Gal & 2M++ C15 & Fixed & \bfullchieff & 0.0 \\
Gal & 2M++/SDSS & Fixed & \bsdsschieff & \bsdssdelchieff \\
Gal & 2M++/SDSS$_\textrm{ilos}$ & Fixed & \bsdssdarcychieff & \bsdssdarcydelchieff \\
Gal & 2M++/SDSS$_\textrm{ilos}$ & Variable & \bsdssdarcyvarchieff & \bsdssdarcyvardelchieff \\
Gal & 2MRS & Fixed & \btmrschieff & \btmrsdelchieff \\
Gal & 2MRS$_\textrm{ilos}$ & Fixed & \btmrsiloschieff & \btmrsilosdelchieff \\
Gal & 2MRS$_\textrm{ilos}$ & Variable & \btmrsilosvarchieff & \btmrsilosvardelchieff \\
\hline
Gal + T15 Group Avg & 2M++ C15 & Fixed & \bTfullchieff & \bTfulldelchieff \\
Gal + T15 Group Avg & 2M++/SDSS & Fixed & \bTsdsschieff & \bTsdssdelchieff \\
Gal + T15 Group Avg & 2M++/SDSS$_\textrm{ilos}$ & Fixed & \bTsdssdarcychieff & \bTsdssdarcydelchieff \\
Gal + T15 Group Avg & 2M++/SDSS$_\textrm{ilos}$ & Variable & \bTsdssdarcyvarchieff & \bTsdssdarcyvardelchieff \\
Gal + T15 Group Avg & 2MRS & Fixed & \bTtmrschieff & \bTtmrsdelchieff \\
Gal + T15 Group Avg & 2MRS$_\textrm{ilos}$ & Fixed & \bTtmrsiloschieff & \bTtmrsilosdelchieff \\
Gal + T15 Group Avg & 2MRS$_\textrm{ilos}$ & Variable & \bTtmrsilosvarchieff & \bTtmrsilosvardelchieff \\
\enddata
\tablecomments{PV uncertainties are either treated as `Fixed' (250 km~s$^{-1}$) or `Variable' (posterior variance for the integration along the line of sight).}
\end{deluxetable}

\end{document}